\documentclass[10pt]{article}
\pdfoutput=1
\usepackage{jheppub,amsmath,amssymb}
\usepackage{enumerate}
\usepackage{bm}
\usepackage{bbm}
\usepackage{enumitem}
\usepackage[usenames,dvipsnames]{xcolor}
\usepackage{amsmath}
\usepackage{amsfonts}
\usepackage{amssymb}
\usepackage{graphicx}   
\usepackage{hhline}
\usepackage{subfig}
\usepackage{color}
\usepackage{verbatim}
\usepackage{amsmath,amsthm,amssymb}
\usepackage{lscape}
\usepackage{float}
\usepackage{caption}
\usepackage{lscape}
\usepackage{breqn}
\usepackage{multicol}
\usepackage{graphics}
\usepackage{tikz,todonotes}
\usepackage[utf8]{inputenc}
\usepackage{autobreak}
\usepackage{verbatim}
\usetikzlibrary{arrows,positioning,decorations.markings,decorations.pathmorphing,calc}
\pgfdeclarelayer{edgelayer}
\pgfdeclarelayer{nodelayer}
\usetikzlibrary{decorations.pathreplacing}
\pgfsetlayers{edgelayer,nodelayer,main}
\tikzset{none/.style={draw=none}}
\tikzset{new edge style 2/.style={black}}
\tikzset{new style 0/.style={black}}
\tikzset{rednode/.style={draw=none, scale=0.3pt,fill=red,circle, draw}}
\tikzset{redline/.style={line width=0.3mm,red}}
\tikzset{greyE/.style={line width=0.1mm,gray}}
\usepackage[utf8]{inputenc}
\usetikzlibrary{arrows,positioning,decorations.markings,decorations.pathmorphing,calc}

\definecolor{hyperref}{RGB}{026,028,087}
\newcommand{\beq}{\begin{equation}}
\newcommand{\eeq}{\end{equation}}
\newcommand{\bea}{\begin{eqnarray}}
\newcommand{\eea}{\end{eqnarray}}
\def\be{\begin{equation}}
\def\ee{\end{equation}}

\def\beq{\begin{equation}}
\def\eeq{\end{equation}}

\renewcommand{\k}{\vec{k}}

\newcommand{\mpl}{M_{\rm Pl}}
\newcommand{\K}{\mathcal K}
\newcommand{\U}{\mathcal U}

\renewcommand{\L}{\mathcal L}

\def\be{\begin{equation}}
\def\ee{\end{equation}}
\def\ba{\begin{eqnarray}}
\def\ea{\end{eqnarray}}

\def\nn{\nonumber}

\usepackage[normalem]{ulem}

\def\ba{\begin{eqnarray}}
\def\ea{\end{eqnarray}}

\def\L{\mathcal{L}}
\def\K{\mathcal{K}}

\def\stu{St\"uckelberg }

\def\({\left(}
\def\){\right)}

\def\k{\kappa}
\def\mpl{M_{\rm Pl}}

\begin{document}

\title{Massive Gravity from Double Copy}

\author[a]{Arshia Momeni,}
\author[a]{Justinas Rumbutis,}
\author[a,b]{Andrew J. Tolley}

\affiliation[a]{Theoretical Physics, Blackett Laboratory, Imperial College, London, SW7 2AZ, U.K.}
\affiliation[b]{CERCA, Department of Physics, Case Western Reserve University, 10900 Euclid Ave, Cleveland, OH 44106, USA}

\emailAdd{arshia.momeni17@imperial.ac.uk}
\emailAdd{j.rumbutis18@imperial.ac.uk}
\emailAdd{a.tolley@imperial.ac.uk}

\abstract{We consider the double copy of massive Yang-Mills theory in four dimensions, whose decoupling limit is a nonlinear sigma model. The latter may be regarded as the leading terms in the low energy effective theory of a heavy Higgs model, in which the Higgs has been integrated out. The obtained double copy effective field theory contains a massive spin-2, massive spin-1 and a massive spin-0 field, and we construct explicitly its interacting Lagrangian up to fourth order in fields.  We find that up to this order, the spin-2 self interactions match those of the dRGT massive gravity theory, and that all the interactions are consistent with a $\Lambda_3= (m^2 \mpl)^{1/3}$ cutoff. We construct explicitly the $\Lambda_3$ decoupling limit of this theory and show that it is equivalent to a bi-Galileon extension of the standard $\Lambda_3$ massive gravity decoupling limit theory. Although it is known that the double copy of a nonlinear sigma model is a special Galileon, the decoupling limit of massive Yang-Mills theory is a more general Galileon theory. This demonstrates that the decoupling limit and double copy procedures do not commute and we clarify why this is the case in terms of the scaling of their kinematic factors.}

\maketitle

%\newpage
%\tableofcontents
%\setcounter{tocdepth}{1}

%%%%%%%%%%%%%%%%
\section{Introduction}
The Bern-Carrasco Johansson (BCJ) {\it double copy} \cite{Bern:2008qj, Bern:2010ue} is a relation between the scattering amplitudes of two different theories. The BCJ relation, or colour-kinematics duality, states that in a gauge theory, one can always represent kinematic factors of scattering amplitudes so that they satisfy an analogue relation to the gauge group colour factors. Replacing the colour factors by kinematic factors in a given theory, leads to new scattering amplitudes describing other theories.\\

The first and most important example is the relationship between Yang-Mills theory and gravity amplitudes \cite{Bern:2010ue}. The origin of this relation can be understood from the string theory point of view by considering how open and closed string amplitudes are related, and looking at the low energy effective field theories of the two string theories. This is encapsulated by the KLT relations \cite{Kawai:1985xq}. However the `double copy' paradigm has been found to be more general, and there are known examples of extensions of double copy relations between two non-gravitational theories for example non-linear sigma model and DBI or special Galileon theories \cite{Chen:2013fya, Cachazo:2014xea,Hinterbichler:2015pqa,Cheung:2017ems,Du:2016tbc, Chen:2014dfa, Chen:2016zwe, Cheung:2017yef, Cheung:2016prv} as well as extended gravitational relations such as that between super Yang-Mills and supergravity theories \cite{Bern:2019prr,Borsten:2015pla,Anastasiou:2017nsz}. Recently the double copy paradigm was extended for gauge theories with massive matter fields \cite{Bautista:2019tdr,Johansson:2019dnu,Plefka:2019wyg,Bautista:2019evw}. \\

The physical applications of double copy extend beyond calculations of scattering amplitudes in Minkowski spacetime. For example, double copy is used for UV considerations of effective field theories \cite{Carrillo-Gonzalez:2019aao, Low:2019wuv, Carrasco:2019qwr, Carrasco:2019yyn}, efficient gravitational wave calculations \cite{Bern:2019crd, Goldberger:2016iau, Shen:2018ebu, Cheung:2018wkq, Kosower:2018adc, Bern:2019nnu} and relations between classical solutions in different theories (known as classical double copy) \cite{Saotome:2012vy, Monteiro:2014cda, Luna:2015paa, Luna:2016due, White:2016jzc, Cardoso:2016amd, Luna:2016hge, Goldberger:2017frp, Ridgway:2015fdl, De_Smet_2017, Bahjat_Abbas_2017, carrillogonzalez2017classical, Goldberger_2018, Li_2018, Lee_2018, Plefka_2019, Berman_2019, Kim:2019jwm, Goldberger:2019xef, Alawadhi:2019urr, Banerjee:2019saj,Huang:2019cja}. The double copy has been shown to apply for scattering amplitudes around more general backgrounds \cite{Adamo:2017nia,Borsten:2019prq}. \\

In this paper we initiate the application of the double copy paradigm to the scattering amplitudes of massive Yang-Mills theory, {\it i.e.} the low energy effective field theory of Yang-Mills coupled to a heavy Higgs field (with the Higgs integrated out) which spontaneously breaks the gauge symmetry in a way that all of the gauge bosons acquire the same mass. On the gauge theory side, the act of spontaneously breaking symmetries is well understood and is a major component of the standard model. Double copy of gauge theories with spontaneously broken gauge symmetries have been studied in \cite{Chiodaroli:2015rdg, Chiodaroli:2017ehv, Chiodaroli:2018dbu}, however the case where both of the copies of gauge theory have completely broken gauge symmetry ({\it i.e.} with only massive gauge bosons) has not been explored. On the gravitational side, the broken gauge symmetries (by virtue of the mass for the bosons) imply if the double copy procedure is still valid, broken diffeomorphism symmetries. The latter are in the purview of massive gravity theories\footnote{See for example \cite{deRham:2014zqa} for an extensive review of recent work in this area.}, and so we may naturally expect massive gravity in some form to arise from the double copy procedure. \\

Since a massive spin-1 particle has 3 degrees of freedom in four dimensions, the double copy theory contains 9 propagating states, which decompose into a single massive spin-2 particle, a single massive spin-1 particle and a massive scalar. The interactions of massive spin-2 particles are well known to be highly constrained. Generic interactions are expected to lead to a breakdown of perturbative unitarity at the $\Lambda_5 = (m^4 \mpl)^{1/5}$ scale \cite{ArkaniHamed:2002sp}, where $m$ is the spin-2 mass. Special tunings can be made that raise this scale to the $\Lambda_3= (m^2 \mpl)^{1/3}$ scale which is the highest possible scale in four dimensions \cite{deRham:2010ik,deRham:2010kj}. An explicit nonlinear effective theory exhibiting this scale is the so-called ghost-free massive gravity or de Rham-Gabadadze-Tolley (dRGT) model \cite{deRham:2010kj}. \\

Remarkably, we find that the double copy paradigm automatically leads to a theory in which the interactions of the massive spin-2 field are described by the dRGT massive gravity \cite{deRham:2010kj}, at least to quartic order. In fact we will find that the free coefficients in the dRGT Lagrangian are fixed by the double copy prescription to this order. We further find that the interactions of the additional spin-1 and spin-0 states are also at the scale $\Lambda_3$ strongly suggesting that this is the controlling scale of the EFT at all orders. Since massive Yang-Mills is itself an EFT with the highest possible cutoff for a coloured spin-1 particle, namely $\Lambda = m/g$, we may regard this as a natural double copy relation between two highest cutoff effective theories. \\

This connection is emphasized when we recognize that the leading helicity-0 interactions of a massive graviton are dominated in the decoupling limit (defined by taking $m \rightarrow 0$ for fixed $\Lambda_3$) by the double copy of the leading helicity-zero interactions of the massive spin-1 gluon. Since the decoupling limit of massive Yang-Mills is a nonlinear sigma model, as encoded in the Goldstone equivalence theorem, we may reasonably expect that the interactions for the helicity-0 spin-2 states are determined by the double copy of the nonlinear sigma model\footnote{This was for example explicitly proposed in \cite{Carrillo_Gonz_lez_2018}.}. It is known that double copy of a non-linear sigma model is the special Galileon \cite{Cachazo:2014xea,Hinterbichler:2015pqa,Cheung:2017ems,Cheung:2017yef, Cheung:2016prv} and that the decoupling limits of massive gravity theories are also Galileon-like theories \cite{ArkaniHamed:2002sp,deRham:2010ik,deRham:2010kj,Ondo:2013wka}. However, the latter are nevertheless more complicated and include in particular non-trivial vector scalar interactions that survive even in the decoupling limit \cite{Ondo:2013wka,Gabadadze:2013ria}. Even projecting onto the scalar sector, the massive gravity decoupling limit is not equivalent to a special Galileon, and so we find that the decoupling limit procedure does not commute with the double copy procedure. \\ 

The origin of this is that there are terms needed in the kinematic factors to satisfy colour-kinematics duality that are singular in the decoupling limit but nevertheless cancel out of the gauge amplitudes. However when we construct the gravity amplitudes by squaring these kinematic factors, they no longer cancel and give additional non-zero contributions that are finite in the decoupling limit. To be precise, the kinematic factors which satisfy colour-kinematics duality $n_s+n_t+n_u=0$ take the form
\be\label{kinematic}
n_s = \frac{s-m^2}{m^3} \Sigma(s,t,u) + \frac{1}{m^2} \hat n_s, \quad n_t = \frac{t-m^2}{m^3} \Sigma(s,t,u) + \frac{1}{m^2} \hat n_t, \quad n_u = \frac{u-m^2}{m^3} \Sigma(s,t,u) + \frac{1}{m^2} \hat n_u \, ,
\ee
where $\Sigma(s,t,u)$ (triple crossing symmetric) and $\hat n_i$ are finite as $m \rightarrow 0$. Here $\Sigma$ arises in a manner similar to the generalized gauge transformations in the massless case, a fact which is crucial to understanding why its contribution is finite. The explicit expressions for $\Sigma$ and $\hat n_i$ are given in Eqs. \eqref{eq: Sigma}, \eqref{eq: nshat}, \eqref{eq: nthat} and \eqref{eq: nuhat}. Since in the massive case $s+t+u = 4m^2$ we have $\hat n_s+\hat n_t+\hat n_u=-m \Sigma$ and so in the limit $m \rightarrow 0$, $\hat n_i$ by themselves satisfy colour-kinematics duality. The $1/m^3$ behaviour in $n_i$ comes from helicity $0,0,0,\pm 1$ interactions since the polarization tensor for a massive helicity-0 gluon scales as $1/m$  but that for helicity-1 is finite as $m \rightarrow 0$. The term $\Sigma$ cancels out of the gauge theory amplitudes
\be
    A^{\text{mYM}}_{4}=g^2 \left(\frac{c_s n_s}{s-m^2}+\frac{c_t n_t}{t-m^2}+\frac{c_u n_u}{u-m^2}\right) =  \frac{1}{\Lambda^2}\left(\frac{c_s \hat n_s}{s-m^2}+\frac{c_t \hat n_t}{t-m^2}+\frac{c_u \hat n_u}{u-m^2}\right),
\ee
by virtue of the colour relation $c_s+c_t+c_u=0$, demonstrating the natural decoupling limit scaling. \\

 By contrast, when we square to construct the gravity amplitudes, $\Sigma$ survives as a contact term. For instance the naive leading $1/m^6$ term enters in the gravity amplitudes in the combination
\be
\frac{1}{\mpl^2} \(\frac{n_sn_s'}{s-m^2}+\frac{n_tn_t'}{t-m^2}+\frac{n_un_u'}{u-m^2} \)\sim \frac{\Sigma \Sigma'}{\mpl^2 m^6} \((s-m^2) + (t-m^2) + (u-m^2) \) +\dots \sim \frac{\Sigma \Sigma'}{\Lambda_3^6} +\dots \, ,
\ee
and hence it contributes at the $\Lambda_3$ scale.  Specifically this will show up as a non-zero spin-2, helicity $0,0,0,\pm 2$ interaction. Similarly the naive $1/m^5$ term is suppressed by virtue of the kinematic relation $\hat n_s+\hat n_t+\hat n_u=-m \Sigma$ and we have in full as an exact statement
\be\label{eq:gravityDL}
\frac{1}{\mpl^2} \(\frac{n_sn_s'}{s-m^2}+\frac{n_tn_t'}{t-m^2}+\frac{n_un_u'}{u-m^2} \)= \frac{-\Sigma \Sigma'}{\Lambda_3^6} +\frac{1}{\Lambda_3^6} \(\frac{\hat n_s \hat n_s'}{s-m^2}+\frac{\hat n_t \hat n_t'}{t-m^2}+\frac{\hat n_u \hat n_u'}{u-m^2} \) \, .
\ee
Since $\Sigma$ does not contribute to the gauge theory amplitudes, first taking the decoupling limit of them (giving a non-linear sigma model) and performing the double copy procedure (giving a special Galileon) will lead to a different result in which the $\frac{\Sigma \Sigma'}{\Lambda_3^6} $ term is absent\footnote{It is of course technically true that if we only compute amplitudes in which the spin-1 helicity-1 polarizations are set to zero, then $\Sigma=\Sigma'=0$ and we will recover the special Galileon amplitudes in the decoupling limit. But this is an inconsistent procedure from the point of view of the gravity theory, and has no relation to the massive gravity theory whose decoupling limit is a special Galileon.
There may however exist an extension of the recipe along the lines discussed in \cite{Johansson:2014zca,Luna:2017dtq,Bern:2019nnu,Bern:2019crd} which allows for a consistent removal of additional degrees of freedom.
}. The kinematic factors inferred from the decoupling limit $\hat n_i(m=0)$ will necessarily be finite in the decoupling limit, and these do not correspond to the decoupling limit of the above kinematic factors \eqref{kinematic} which are singular. Indeed in the decoupling limit, the gauge theory kinematic factors come purely from helicity-0 gluons by the Goldstone equivalence theorem.\\

It is worth noting that if we give up strict colour-kinematics duality in the massive case, then an acceptable choice of kinematic factors that reproduce the gauge theory amplitudes are $\tilde n_i= \hat n_i/m^2$. However they no longer sum to zero. Using these in a double copy prescription will give a gravity amplitude given by the second term on the RHS of \eqref{eq:gravityDL}, whose decoupling limit correctly reproduces the special Galileon. However, since $\sum_i \tilde n_i \neq 0$ we have no reason to trust that the double copy prescription is meaningful in this context. Indeed, there is no clear recipe to generalize this to higher amplitudes. It is for this reason that throughout this paper we assume that the colour-kinematics duality holds in tact in the massive case in the same manner as the massless.\\

The paper is organised as follows: first we briefly introduce massive Yang-Mills in section \ref{sec:ym} and dRGT massive gravity theories in section \ref{sec:mgr}, then describe the double copy prescription and give the action obtained from squaring massive Yang-Mills in section \ref{sec:main}. In particular we find that the colour-kinematics duality holds for 2-2 scattering amplitudes and the resulting theory has $\Lambda_3=(m^2\mpl)^{1/3}$ cutoff scale which is known to be the highest possible cutoff for massive spin-2 fields \cite{ArkaniHamed:2002sp}. Having determined the gravity Lagrangian up to quartic order, we specify the decoupling limit in section \ref{sec:DL} and clarify its inequivalence to a special Galileon. The precise quartic interactions are given in Appendix \ref{sec:contact} and our conventions are give in Appendix \ref{sec:conv}. Appendix \ref{sec:dual} contains a brief explanation of why giving a mass to a two form potential (which arises naturally in the massless double copy story) is equivalent to a massive spin-1 Proca theory, as we find the latter formulation more useful in constructing the interacting Lagrangian. In Appendix \ref{sec:dl_mg} we complement section \ref{sec:DL} and give the explicit decoupling limit of the gravity amplitudes, while in Appendix \ref{sec:dl ym} we do the same for the Yang-Mills amplitudes and clarify why the double copy procedure does not commute with the decoupling limit.

\subsection*{\bf Note added}

In preparing this work for submission we became aware of results obtained by Laura Johnson, Callum Jones and Shruti Paranjape which also reproduce the quartic double copy interactions \cite{Johnson:2020pny}.

%%%%%%%%%%%%%%%%

\section{Massive Yang-Mills}\label{sec:ym}
The action of massive Yang-Mills theory comes from the low energy effective action of Yang-Mills theory with a Higgs field in which the Higgs particles are integrated out. We consider the gauge symmetry to be broken in such a way that all of the gauge bosons acquire the same mass, $m$. Then the leading terms in the effective Lagrangian in unitary gauge are as follows:

\begin{equation}\label{eq:lagr mYM}
    \L_{mYM}=-\frac{1}{4} \text{tr}(F_{\mu\nu}F^{\mu\nu})-\frac{1}{2} m^2 \text{tr}(A_{\mu} A^{\mu}),
\end{equation}
where $g$ is the coupling constant. This is the simplest unitary gauge Lagrangian which can describe a massive coloured spin-1 particle. Since the resulting theory is not renormalizable, it should be understood as an effective theory, and to this Lagrangian we may add an infinite number of interactions. For instance, we may further consider a quartic interaction  $ \text{tr}(A_{\mu} A^{\mu})^2$. The structure of the effective Lagrangian is best understood by reintroducing \stu fields (Goldstone modes) by replacing
\be
A_{\mu} \rightarrow \frac{\sqrt{2} i}{g} V(x)^{-1} D_{\mu} V(x)
\ee
where $D_{\mu} = \partial_{\mu} -\frac{i g}{\sqrt{2}} A_{\mu}$ is the covariant derivative and $V(x)=\exp\left[ \frac{i}{\sqrt{2} \Lambda} T^a \phi^a(x)\right]$ where $\phi^a(x)$ are the \stu fields, so that the gauge invariant form of the Lagrangian is
\begin{equation}\label{eq:lagr mYM1}
    \L_{mYM}=-\frac{1}{4} \text{tr}(F_{\mu\nu}F^{\mu\nu})- \Lambda^2 \text{tr}(D_{\mu} V D^{\mu} V^{-1}),
\end{equation}
where $\Lambda = m/g$. This Lagrangian is manifestly gauge invariant under $D_{\mu} \rightarrow U(x)^{-1} D_{\mu} U(x)$ under which the \stu fields transform as $V(x) \rightarrow U(x)^{-1} V(x)$ where $U(x)=\exp\left[ \frac{i}{\sqrt{2} \Lambda} T^a \xi^a(x)\right]$ and $\xi^a(x)$ is the gauge transformation parameter. The unitary gauge Lagrangian is recovered by fixing the gauge $\phi^a =0$. \\

The resulting effective theory has a cutoff of at most $\Lambda=m/g$ which is the Goldstone mode decay constant. Additional interactions in the effective action could further lower this scale, but for now we assume that $\Lambda$ is the controlling scale. Taking the decoupling limit $g \rightarrow 0$ for fixed $\Lambda$ results in a free massless spin-1 theory and an interacting non-linear sigma model
\begin{equation}\label{eq:lagr mYM2}
    \L_{DL}=\lim_{g \rightarrow 0, \Lambda \text{ fixed}} \L_{mYM}=-\frac{1}{4} \sum_a (\partial_{\mu} A^a_{\nu} - \partial_{\nu} A^a_{\mu} )^2 - \Lambda^2 \text{tr}(\partial_{\mu} V \partial^{\mu} V^{-1}).
\end{equation} \\
This encodes straightforwardly the content of the `Goldstone equivalence theorem' that the leading interactions for the helicity-0 modes of the massive spin-1 particle are determined by the effective theory for the Goldsones described by \eqref{eq:lagr mYM2}. From a classical perspective, the form of the Lagrangian \eqref{eq:lagr mYM1} is clearly preferred due to its two derivative nature and it is for the reason that we will focus on the tree level amplitudes derived from this form in what follows. Were we to include additional unitary gauge interactions such as $ \text{tr}(A_{\mu} A^{\mu})^2$, etc. it is transparent in the \stu formulation that these correspond to higher order operators, and they are expected to be suppressed by the scale $\Lambda$. In the decoupling limit, these extensions just correspond to the addition of further irrelevant operators to the nonlinear sigma model Lagrangian, which have been considered in the double copy context for example in \cite{Carrillo-Gonzalez:2019aao, Low:2019wuv, Carrasco:2019qwr, Carrasco:2019yyn}. \\

These tree amplitudes are however most conveniently computed in unitary gauge \eqref{eq:lagr mYM}. This is because the off-shell vertices for massive Yang-Mills are identical to their massless counterparts, and the only difference is the massless propagator is replaced by the massive one with structure
\be
\frac{-i \hat \eta_{\mu\nu}} {p^2+m^2} \, ,
\ee
where $ \hat \eta_{\mu\nu} = \eta_{\mu\nu} + p_{\mu} p_{\nu}/m^2$. Our goal is to follow as closely as possible the double copy paradigm for massless Yang-Mills theory \cite{Bern:2010ue} and with this in mind we express the tree level $n$-point scattering amplitudes of this theory as:
\begin{equation}\label{eq:An}
A_n=g^{n-2}\sum_{i}\frac{c_i n_i}{\prod_{\alpha_{i}}(-p_{\alpha_{i}}^2-m^2)},    
\end{equation}
where $c_i$ are colour factors {\it i.e.} products of the structure constants of the gauge group, $n_i$ are the kinematic factors, $i$ labels distinct Feynman graphs and $\alpha_{i}$ labels all internal propagators in a given graph. The only difference between this and the standard double copy is the replacement of massless propagators $p_{\alpha_{i}}^2$ by massive $p_{\alpha_{i}}^2+m^2$. The resulting kinematic factors $n_i$ are not the same as those that arise in the massless case since they absorb the information from the massive polarization structure encoded in $\hat \eta_{\mu\nu}$, and furthermore the on-shell external momenta now satisfy $p^2_i = -m^2$. Given this it is not automatic that the colour-kinematics duality still holds. We will nevertheless show that it continues to hold up to quartic order.

\subsection{Three-point Amplitude}
In terms of polarization and momentum vectors the three-point on-shell vertex for massive Yang-Mills is exactly same as that of massless Yang-Mills:\footnote{All of our scattering amplitudes are given as the momentum space delta function stripped amplitudes of $\langle \{ k_f\} | \hat S-\hat 1 | \{ k_i \} \rangle$. {\it i.e.} we forgo the introduction of an $i$ as in $\hat S = \hat 1 + i \hat T$.}
\be\label{eq:a3}
A_3(1^a,2^b,3^c)=\sqrt{2} g f_{abc}(-\epsilon_{1}\cdot\epsilon_{2}\;\epsilon_{3}\cdot p_{1}+\epsilon_{1}\cdot\epsilon_{3}\;\epsilon_{2}\cdot p_{1}-\epsilon_{1}\cdot p_{2}\;\epsilon_{2}\cdot\epsilon_{3}).
\ee
The difference is that now the on-shell momenta satisfy $p_i^2=-m^2$ and there are 3 possible polarization states. Our conventions for these are given in Appendix \ref{sec:conv}.

\subsection{Four-point Amplitude}
We express the four-point amplitude in the form given in Eq. \eqref{eq:An} by defining the colour factors to be:
\begin{align}
&c_s=f_{abe}f_{cde}\\
&c_t=f_{cae}f_{bde}\\
&c_u=f_{bce}f_{ade}.
\end{align}
so that
\be
A_4(1^a,2^b,3^c,4^d) = g^2 \left( \frac{c_s n_s}{s-m^2} + \frac{c_t n_t}{t-m^2} +\frac{c_u n_u}{u-m^2} \right) \, ,
\ee
where the kinematic factors are
\begin{align}\label{eq:ns}
\begin{autobreak}
 n_s= 
-\frac{i}{2} (\epsilon_1\cdot\epsilon_2 p_1\cdot\epsilon_3 p_1\cdot\epsilon_4
+4 \epsilon_2\cdot\epsilon_4 p_1\cdot\epsilon_3 p_2\cdot\epsilon_1
-2 \epsilon_2\cdot\epsilon_3 p_1\cdot\epsilon_4 p_2\cdot\epsilon_1
+3 \epsilon_1\cdot\epsilon_2 p_1\cdot\epsilon_4 p_2\cdot\epsilon_3
+4 \epsilon_2\cdot\epsilon_4 p_2\cdot\epsilon_1 p_2\cdot\epsilon_3
-4 \epsilon_1\cdot\epsilon_4 p_1\cdot\epsilon_2 (p_1\cdot\epsilon_3
+p_2\cdot\epsilon_3)
-3 \epsilon_1\cdot\epsilon_2 p_1\cdot\epsilon_3 p_2\cdot\epsilon_4
-2 \epsilon_2\cdot\epsilon_3 p_2\cdot\epsilon_1 p_2\cdot\epsilon_4
-\epsilon_1\cdot\epsilon_2 p_2\cdot\epsilon_3 p_2\cdot\epsilon_4
+4 \epsilon_3\cdot\epsilon_4 p_1\cdot\epsilon_2 p_3\cdot\epsilon_1
-4 \epsilon_3\cdot\epsilon_4 p_2\cdot\epsilon_1 p_3\cdot\epsilon_2
+2 \epsilon_1\cdot\epsilon_3 p_1\cdot\epsilon_2 (p_1\cdot\epsilon_4
+p_2\cdot\epsilon_4
-p_3\cdot\epsilon_4)
+\epsilon_1\cdot\epsilon_2 p_1\cdot\epsilon_3 p_3\cdot\epsilon_4
+2 \epsilon_2\cdot\epsilon_3 p_2\cdot\epsilon_1 p_3\cdot\epsilon_4
-\epsilon_1\cdot\epsilon_2 p_2\cdot\epsilon_3 p_3\cdot\epsilon_4
+\epsilon_1\cdot\epsilon_4 \epsilon_2\cdot\epsilon_3  (m^2
-s)
+\epsilon_1\cdot\epsilon_3 \epsilon_2\cdot\epsilon_4  (
-m^2
+s)
+\epsilon_1\cdot\epsilon_2 \epsilon_3\cdot\epsilon_4 t
-\epsilon_1\cdot\epsilon_2 \epsilon_3\cdot\epsilon_4 u),
\end{autobreak}
\end{align}
\begin{align}\label{eq:nt}
\begin{autobreak}
n_t= 
\frac{i}{2} (\epsilon_1\cdot\epsilon_3 p_1\cdot\epsilon_2 p_1\cdot\epsilon_4
+4 \epsilon_2\cdot\epsilon_4 p_1\cdot\epsilon_3 p_2\cdot\epsilon_1
+\epsilon_1\cdot\epsilon_3 p_1\cdot\epsilon_2 p_2\cdot\epsilon_4
+4 \epsilon_3\cdot\epsilon_4 p_1\cdot\epsilon_2 p_3\cdot\epsilon_1
-2 \epsilon_2\cdot\epsilon_3 p_1\cdot\epsilon_4 p_3\cdot\epsilon_1
-4 \epsilon_2\cdot\epsilon_4 p_2\cdot\epsilon_3 p_3\cdot\epsilon_1
+2 \epsilon_2\cdot\epsilon_3 p_2\cdot\epsilon_4 p_3\cdot\epsilon_1
+3 \epsilon_1\cdot\epsilon_3 p_1\cdot\epsilon_4 p_3\cdot\epsilon_2
-\epsilon_1\cdot\epsilon_3 p_2\cdot\epsilon_4 p_3\cdot\epsilon_2
+4 \epsilon_3\cdot\epsilon_4 p_3\cdot\epsilon_1 p_3\cdot\epsilon_2
-4 \epsilon_1\cdot\epsilon_4 p_1\cdot\epsilon_3 (p_1\cdot\epsilon_2
+p_3\cdot\epsilon_2)
-3 \epsilon_1\cdot\epsilon_3 p_1\cdot\epsilon_2 p_3\cdot\epsilon_4
-2 \epsilon_2\cdot\epsilon_3 p_3\cdot\epsilon_1 p_3\cdot\epsilon_4
-\epsilon_1\cdot\epsilon_3 p_3\cdot\epsilon_2 p_3\cdot\epsilon_4
+2 \epsilon_1\cdot\epsilon_2 p_1\cdot\epsilon_3 (p_1\cdot\epsilon_4
-p_2\cdot\epsilon_4
+p_3\cdot\epsilon_4)
+\epsilon_1\cdot\epsilon_3 \epsilon_2\cdot\epsilon_4 s
+\epsilon_1\cdot\epsilon_4 \epsilon_2\cdot\epsilon_3  (m^2-t)
+\epsilon_1\cdot\epsilon_2 \epsilon_3\cdot\epsilon_4  (-m^2+t)
-\epsilon_1\cdot\epsilon_3 \epsilon_2\cdot\epsilon_4 u),
\end{autobreak}
\end{align}
\begin{align}\label{eq:nu}
\begin{split}
&n_u= -\frac{i}{2} (4 \epsilon_1\cdot\epsilon_4 p_1\cdot\epsilon_2 p_2\cdot\epsilon_3-4 \epsilon_2\cdot\epsilon_4 p_2\cdot\epsilon_1 p_2\cdot\epsilon_3-4 \epsilon_2\cdot\epsilon_4 p_2\cdot\epsilon_3 p_3\cdot\epsilon_1+4 \epsilon_2\cdot\epsilon_3 p_2\cdot\epsilon_4 p_3\cdot\epsilon_1\\&-4 \epsilon_1\cdot\epsilon_4 p_1\cdot\epsilon_3 p_3\cdot\epsilon_2+4 \epsilon_3\cdot\epsilon_4 p_2\cdot\epsilon_1 p_3\cdot\epsilon_2+4 \epsilon_3\cdot\epsilon_4 p_3\cdot\epsilon_1 p_3\cdot\epsilon_2-4 \epsilon_2\cdot\epsilon_3 p_2\cdot\epsilon_1 p_3\cdot\epsilon_4\\&+4 \epsilon_1\cdot\epsilon_2 p_2\cdot\epsilon_3 (p_2\cdot\epsilon_4+p_3\cdot\epsilon_4)-4 \epsilon_1\cdot\epsilon_3 p_3\cdot\epsilon_2 (p_2\cdot\epsilon_4+p_3\cdot\epsilon_4)+\epsilon_1\cdot\epsilon_4 \epsilon_2\cdot\epsilon_3 s-\epsilon_1\cdot\epsilon_4 \epsilon_2\cdot\epsilon_3 t\\&+\epsilon_1\cdot\epsilon_3 \epsilon_2\cdot\epsilon_4 (m^2-u)+\epsilon_1\cdot\epsilon_2 \epsilon_3\cdot\epsilon_4 (-m^2+u)),
\end{split}
\end{align}
where the Mandelstam variables are defined as standard:
\be
  s=-(p_1+p_2)^2, \quad t=-(p_1+p_3)^2, \quad u=-(p_1+p_4)^2,
\ee
with all incoming momenta. These expressions for kinematic factors are very similar to those obtained from massless Yang-Mills theory but there are two differences: the relation between Mandelstam variables is now $s+t+u=4m^2$ rather than $s+t+u=0$ and the locations of the poles now are at $s,t,u=m^2$. Because of that the terms coming from quartic Yang-Mills vertex now have to be multiplied by $s-m^2$, $t-m^2$ and $u-m^2$ in order to recast the amplitude into the form \eqref{eq:An}.\\ 

In general, kinematic factors of a given scattering amplitude are not unique. They are not invariant under field redefinitions. However in massless Yang-Mills theory for any choice of kinematic factors of four-point amplitude, the colour-kinematics duality, $c_s+c_t+c_u=0 \rightarrow n_s+n_t+n_u=0$, is satisfied \cite{Cheung:2017pzi}. In our case of massive Yang-Mills theory, it is not immediately clear whether this is still true. However, explicit calculation shows that our colour and kinematic factors (directly calculated from usual Feynman rules) in \eqref{eq:ns},\eqref{eq:nt} and \eqref{eq:nu} still obey $n_s+n_t+n_u\propto p_4\cdot\epsilon_4=0$ and $c_s+c_t+c_u=0$.
The fact that this still holds for the massive theory can be understood by noticing that the only difference between massive and massless kinematic factors is coming from the terms proportional to $m^2$ in \eqref{eq:ns}, \eqref{eq:nt} and \eqref{eq:nu} (in fact we do not need to use the relation between $s$, $t$ and $u$ here). It is easy to see that these six terms add to zero, therefore the value of $n_s+n_t+n_u$ is the same for massless and massive theory and colour-kinematics duality for four-point amplitude still holds in the massive case.

\section{dRGT Massive Gravity}\label{sec:mgr}
In the dRGT theory of massive gravity, the diffeomorphism symmetry is broken by the non-dynamical reference metric, $f_{\mu\nu}$, which appears in the action. It can be written in unitary gauge in terms of the variables \cite{deRham:2010kj}
\begin{equation}
\mathcal K^\mu_\nu(f,g)=\delta^\mu_\nu-\left(\sqrt{g^{-1}f}\right)^\mu_\nu\;.
\end{equation}
This unusual square root metric structure is what is needed to build a $\Lambda_3$ effective theory as it has a straightforward decoupling limit as we shall see in section \ref{sec:DL}. The full dRGT Lagrangian for a single spin-2 field can then be constructed in unitary gauge as \cite{deRham:2014zqa}
\begin{equation}\label{act0}
\L=\frac{M_{\text{Pl}}^{2}}{2}\sqrt{-g}R+\frac{m^2M_{\text{Pl}}^{2}}{4}\sqrt{-g}\,\sum_{n=0}^4\kappa_n\,\U_n\left[\K\right]\end{equation}
where we set $\kappa_0=\kappa_1=0$ and $\kappa_2$ = 1 and the terms in the potential are defined as
\begin{align}
&\U_2(\mathcal K)=2\left([\K]^2-[\K^2]\right)\;,\\
&\U_3(\K)=[\K]^3-3[\K][\K^2]+2[\K^3]\;,\\
&\U_4(\K)=[\K]^4-6[\K^2][\K]^2+8[\K^3][\K]+3[\K^2]^2-6[\K^4]\;.
\end{align}
The squared brackets denote the traces, and the two coefficients $\kappa_3\,,\kappa_4$ are the free parameters of the theory together with the graviton mass $m^2$. The potential terms can be written in terms of the flat space Levi-Civita tensor\footnote{We use Euclidean coventions so that for flat spacetime $\varepsilon_{0123}=\varepsilon^{0123}=1$, i.e. in the Lorentzian $\varepsilon_{\mu\nu\alpha\beta}=-\eta_{\mu \mu'}\eta_{\nu \nu'}\eta_{\alpha \alpha'}\eta_{\beta \beta'}\varepsilon^{\mu'\nu'\alpha'\beta'}$. As long as we are clear that we use one of them with all indices up and the other with all indices down together with $\varepsilon_{i_1\dots i_ki_{k+1}\dots i_d}\varepsilon^{i_1\dots i_kj_{k+1}\dots j_d}=k!\delta_{i_{k+1}\dots i_d}^{j_{k+1}\dots j_d}$ with the generalized Kronecker delta expressed as a determinant of a matrix built out of $\delta$'s.}
\begin{equation}\label{mass1}
\begin{split}
&\U_2(\K)=\varepsilon_{\mu\nu\alpha\beta}\varepsilon^{\mu\nu\alpha'\beta'}\K^{\alpha}_{\alpha'}\K^{\beta}_{\beta'}\,, \\
&\U_3(\K)=\varepsilon_{\mu\nu\alpha\beta}\varepsilon^{\mu\nu'\alpha'\beta'}\K^{\nu}_{\nu'}\K^{\alpha}_{\alpha'}\K^{\beta}_{\beta'}\,,\\
&\U_4(\K)=\varepsilon_{\mu\nu\alpha\beta}\varepsilon^{\mu'\nu'\alpha'\beta'}\K^{\mu}_{\mu'}\K^{\nu}_{\nu'}\K^{\alpha}_{\alpha'}\K^{\beta}_{\beta'}\;.
\end{split}
\end{equation}
In this paper we will consider $f_{\mu\nu}$ to be Minkowski metric, $\eta_{\mu\nu}$ as we shall be largely concerned with scattering amplitudes in Minkowski spacetime. The terms in \eqref{act0} are the unique interactions which lead to second order equations of motion for all 5 propagating degrees of freedom. However from the EFT perspective it is natural to view them as the leading terms in an EFT expansion, controlled by the scale $\Lambda_3$.  Possible higher derivative operators will arise schematically as 
\be
\Delta \L = \Lambda_3^4 \sqrt{-g} \, F[g_{\mu\nu}, K_{\mu \nu}, \frac{\nabla_{\mu}}{\Lambda_3} , \mpl R_{\mu \nu \rho \sigma}] \, .
\ee
where $F$ denotes the sum of all diffeomorphism invariant scalar operators\footnote{All breaking of diffeomorphism invariance can be captured by the tensor $\K_{\mu\nu}$, hence all terms in the Lagrangian are diffeomorphism invariant when $\K_{\mu\nu}$ itself is viewed to transform as a tensor.} constructed out of its arguments with dimensionless Wilson coefficients.\\

Just as for a massive Yang-Mills field we can write this same Lagrangian in a manifestly covariant way via the introduction of \stu fields. Since \eqref{act0} is written an a manner in which it would be manifestly covariant if $K^\mu_\nu(f,g)$ itself transforms as a tensor, then this tells us how to introduce \stu fields. Since the only part of $K^\mu_\nu(f,g)$ that does not transform appropriately as a tensor is the reference metric $f_{\mu\nu} = \eta_{\mu\nu}$, it is sufficient to write this metric in an arbitrary coordinate system
\be
f_{\mu\nu} \rightarrow  \partial_{\mu} \Phi^A \partial_{\nu} \Phi^B \eta_{AB} \, .
\ee
The four diffeomorphism scalars $\Phi^A(x)$ may then be split as $\Phi^A(x)=x^A + \pi^A(x)$. The $\pi^A(x)$ are then the \stu fields we need to reintroduce manifest diffeomorphism invariance and play the analogue of the $\phi^a(x)$ in $V(x)$ \eqref{eq:lagr mYM1}, so that unitary gauge is $\pi^A(x)=0$. We will make explicit use of this decomposition in section \ref{sec:DL}. For the purposes of calculating scattering amplitudes it is sufficient to work with the unitary gauge Lagrangian.

\subsection{Three-point Amplitude}
The three-point amplitude in dRGT massive gravity is as follows:
\be
M_{3}\propto\epsilon_{1\mu\nu}\epsilon_{2\rho\sigma}\epsilon_{3\alpha\beta}\Gamma^{\mu\nu\rho\sigma\alpha\beta}_3
\ee
where $\Gamma_3$ is the cubic vertex from Einstein-Hilbert term plus the cubic potential term $\U_3(\K)$. It is expressed as follows:
\begin{align}
\begin{split}
 M_3=&i\kappa\Big((\epsilon^{\mu \nu }_{1}\epsilon _{3\mu \nu }\epsilon _{2\alpha \beta }p^{\alpha }_{1}p^{\beta }_{1} +2\epsilon _{1 \mu \nu }\epsilon ^{\mu \alpha }_{2}\epsilon^{\nu }_{3\beta} p_{1\alpha }p_{2\beta }+\text{cyclic permutations of 1,2,3})\\&+ \frac{3}{2}  (1+\kappa_3) \epsilon_1^{\mu \nu }\epsilon_{2 \nu\alpha }\epsilon_{3\mu}^{\alpha}  m^2\Big),
\end{split}
\end{align}
where the coupling constant $\kappa=2/\mpl$. The first term is already proportional to the square of Yang-Mills three-point colour-stripped amplitude if we write the polarization tensors as products of two spin-1 polarization vectors, $(\epsilon^i)_{\mu\nu}=(\epsilon^i)_{\mu}(\epsilon^i)_{\nu}$ \footnote{Note that only polarization tensors for helicity $\pm2$ can be written as $(\epsilon^i)_{\mu\nu}=(\epsilon^i)_{\mu}(\epsilon^i)_{\nu}$, for helicities $\pm1,0$ we need to sum over the products of different helicities weighted by Clebsch–Gordan coefficients $\epsilon^{\lambda}_{\mu\nu}=\sum_{\lambda'\lambda''}C_{\lambda'\lambda''}^{\lambda}\epsilon^{\lambda'}_{\mu}\epsilon^{\lambda''}_{\nu}$.}, $M_3$. Therefore, in order for double copy to work we need to choose $\kappa_3$ such that the second term vanishes, {\it i.e.} $\kappa_3=-1$. We see that already at cubic level the double copy construction picks a particular one parameter ($\kappa_4$) subset of theories from 2-parameter family of massive gravity theories.

\section{Double Copy of Massive Yang-Mills}\label{sec:main}
\subsection{Degrees of Freedom}\label{sec:dof}
In the double copy construction the asymptotic states in the gravitational theory are identified with the tensor products of gauge theory asymptotic states, ignoring their colour indices. For example, the double copy of pure Yang-Mills theory gives the following states:
\begin{equation} \label{eq:A times A}
    A_{\mu}	\otimes A_{\nu}=h_{\mu\nu}\oplus B_{\mu\nu}\oplus \phi,	
\end{equation}
{\it i.e.} we decompose the tensor product of two massless vector representation into irreducible representations of Lorentz group: $h_{\mu\nu}$ is the graviton, $B_{\mu\nu}$ is a massless antisymmetric 2-form field and $\phi$ is a massless scalar field (dilaton). In four dimensions the massless $B_{\mu\nu}$ is dual to a pseudo-scalar, {\it i.e.} axion. In terms of degrees of freedom we have $2 \times 2 = 2+ 1^* + 1$.
\\

In the case of massive Yang-Mills, all the fields in \eqref{eq:A times A} are massive: $h_{\mu\nu}$ is a massive spin-2 field, $B_{\mu\nu}$ is a massive 2-form field which is dual to a massive spin-1 field in four dimensions and $\phi$ is a massive scalar field. In terms of degrees of freedom we now have $3 \times 3 = 5+ 3 + 1$. In this paper we will consider four dimensions and write the action obtained from double copy of massive Yang-Mills in terms of massive spin-2 ($h_{\mu\nu}$), massive spin-1 ($A_{\mu}$) and massive spin-0 ($\phi$) fields. We see that there is an interesting physical difference between the field content of the double copy of massless and massive Yang-Mills theories: in the massless case the $B$ field is a spin-0 field while in massive case it is spin-1.

\subsection{Double Copy Construction of Scattering Amplitudes}
In order to double copy massless Yang-Mills theory the representation for the amplitude in \eqref{eq:An} must satisfy the colour-kinematics duality \cite{Bern:2008qj}, {\it i.e.} whenever three of the colour factors, $c_i$, $c_j$ and $c_k$ are related by the Jacobi identity, $c_i+c_j+c_k=0$, the corresponding kinematic factors must obey the same relation {\it i.e.} $n_i+n_j+n_k=0$. It is conjectured \cite{Bern:2008qj} that it is always possible to choose a representation for the amplitude for which kinematic factors satisfy this by choosing a gauge and performing field redefinitions. In the massive case that is not guaranteed to be true but we have checked that the kinematic factors of four-point amplitude calculated directly from \eqref{eq:lagr mYM} satisfy the colour-kinematics duality.

In the usual double copy procedure, once the correct representation for \eqref{eq:An} is chosen, the colour factors can be replaced with kinematic factors in order to obtain an amplitude of a gravitational theory \cite{Bern:2010ue}. We follow the same procedure and conjecture that the following expression gives an amplitude in a massive gravity theory:
\begin{equation}\label{eq:Mn}
M_n=i\left(\frac{\kappa}{2}\right)^{n-2}\sum_{i}\frac{n_i \tilde n_i}{\prod_{\alpha_{i}}(-p_{\alpha_{i}}^2-m^2)},    
\end{equation}
where $\tilde n_i$ are the kinematic factors of the second massive Yang-Mills. The products of Yang-Mills polarization tensors in $n_i$ and $\tilde n_i$, $\epsilon_{\mu}$ and $\tilde\epsilon_{\nu}$ respectively, are decomposed into polarization tensors of the fields in the gravitational theory. This corresponds to decomposition of a tensor product of two vector representations of the little group (for massive particles in 4d it is $SO(3)$) into irreducible representations. Schematically this is done as follows:
\begin{align}\label{eq:pol h}
    &\epsilon^{((j}_{\mu}\tilde\epsilon^{\k))}\rightarrow \epsilon^{(h)jk}_{\mu\nu}\\ 
    &\epsilon_{\mu}^{[j}\tilde\epsilon_{\nu}^{k]}\rightarrow \epsilon^{(B)jk}_{\mu\nu} ,\label{eq:pol B}\\
    &\epsilon^{j}_{\mu}\tilde\epsilon^{\k}\delta_{jk}\propto\epsilon^{(\phi)}_{\mu\nu}. \label{eq:pol tr}
\end{align}
where $j,k$ are little group indices, $(())$ denotes the symmetric traceless part corresponding to the graviton polarization, $\epsilon^{(h)}$, and the antisymmetric part denoted as $[]$ corresponds to the spin-1 polarization in terms of the $B$ field, $\epsilon^{(B)}$. However instead of working with the massive $B_{\mu\nu}$ field in this paper, we construct the action in terms of the vector field $A_{\mu}$ which is dual to $B_{\mu\nu}$. The dualization procedure is explained in Appendix \ref{sec:dual}. We define the map between $B$ field polarization tensor and $A$ polarization vector to be:
\be\label{eq:pol map}
\epsilon^{(B)}_{\mu\nu}=\frac{i}{\sqrt{2}m}\varepsilon_{\mu\nu\rho\sigma}p^{\rho}\epsilon^{(A)\sigma},
\ee
where $p^{\sigma}$ is the four-momentum of the external state and the factor of $\sqrt{2}$ is required for the correct normalization. The trace part of the tensor product, given in \eqref{eq:pol tr}, is the polarization tensor corresponding to the scalar, $\phi$. As we show in \ref{sec:CG} from explicit calculation in helicity basis we find it to be \be\label{eq:pol phi}
\epsilon^{(\phi)}_{\mu\nu}=\frac{1}{\sqrt{3}}\left(\eta_{\mu\nu}+\frac{p_{\mu}p_{\nu}}{m^2}\right),
\ee
which up to a sign could equally have been fixed by the requirement that it is a tracefull, transverse and normalized.

\subsection{Double Copy of Three-point Amplitudes}
We apply \eqref{eq:Mn} to three-point amplitudes explicitly giving the following relation:
\be
M_3=i\frac{\kappa}{2}A_3\tilde A_3,.
\ee
where the 3 point amplitudes have their structure constants, $f_{abc}$, stripped off.
By substituting \eqref{eq:a3} and \eqref{eq:pol h}, \eqref{eq:pol map} and \eqref{eq:pol phi} we get the following three-point vertices in a gravitational theory:

\begin{align}
    &M_{AAh}=i\frac{\kappa}{2}\left(\frac{3}{2}m^2\epsilon_1^{\mu}\epsilon_2^{\nu}\epsilon_{3\mu\nu}-p_1^{\alpha}p_2^{\beta}\epsilon_1^{\mu}\epsilon_2^{\mu}\epsilon_{3\alpha\beta}+p_1^{\mu}p_2^{\nu}\epsilon_1^{\nu}\epsilon_2^{\alpha}\epsilon_{3\mu\alpha}+p_1^{\mu}p_2^{\nu}\epsilon_1^{\alpha}\epsilon_{2\mu}\epsilon_{3\nu\alpha}\right)\\
    &M_{AA\phi}=-i\frac{\kappa}{8\sqrt{3}}\left(15m^2\epsilon_1^{\mu}\epsilon_{2\mu}+2p_{1\nu}p_{2\mu}\epsilon_1^{\mu}\epsilon_2^{\nu}\right)\\
    &M_{\phi hh}=-i\frac{\sqrt{3}\kappa}{4}m^2\epsilon_{2\mu\nu}\epsilon_{3}^{\mu\nu}\label{eq:Mphihh}\\
    &M_{\phi\phi h}=-i\frac{3\kappa}{4}p_{1\mu}p_{2\nu}\epsilon_3^{\mu\nu}\\
    &M_{\phi\phi\phi}=-i\frac{11\sqrt{3}}{16}\kappa m^2\\
    &M_{hhh}=i\kappa\left((\epsilon ^{\mu \nu }_{1}\epsilon _{3\mu \nu }\epsilon _{2\alpha \beta }p^{\alpha }_{1}p^{\beta }_{1} +2\epsilon _{1\mu \nu }\epsilon ^{\mu \alpha }_{2}\epsilon^{\nu }_{3\beta} p_{1\alpha }p_{2\beta }+\text{cyclic permutations of 1,2,3})\right)\label{eq:Mhhh}
\end{align}

As mentioned before, $M_{hhh}$ matches three graviton amplitude of massive gravity if we choose $\kappa_3=-1$ (or $c_3=1/4$ using the parametrization of the theory as in \cite{deRham:2010ik,Cheung:2016yqr}). The $M_{AAh}$ and $M_{\phi\phi h}$ amplitudes are different from those obtained from vector and scalar kinetic terms minimally coupled to gravity (for example a minimally coupled scalar would give $M_{\phi\phi h}=-i\kappa \epsilon_{3\mu\nu}p_1^{\mu}p_2^{\nu}$. This is expected, since we know theories containing massive spin-2 field do not have diffeomorphism symmetry, and we allow couplings between our fields and the reference metric which in this case is the Minkowski metric. In this way we evade the usual equivalence principle requirements for a massless spin-2 particle. As already mentioned we see that $M_{hhh}$ matches the 3 point amplitude of massive gravity with $\kappa_3=-1$.

\subsection{Double Copy of Four-point Amplitudes}
We start with $hh\rightarrow hh$ amplitude which is calculated using  \eqref{eq:ns}, \eqref{eq:nt}, \eqref{eq:nu}, \eqref{eq:Mn} and \eqref{eq:pol h}. By comparing it with $hh\rightarrow hh$ amplitude calculated using dRGT massive gravity action, $M_{4}^{\text{mGr}}$, we find the following:
\be\label{eq:M4hh}
M_{4}=M_{4}^{\text{mGr}}-i\frac{3}{16}\kappa^2 m^4\left(\frac{ \epsilon_{1\mu\nu}\epsilon_{2}^{\mu\nu}\epsilon_{3\alpha\beta}\epsilon_{4}^{\alpha\beta}}{s-m^2}+\frac{ \epsilon_{1\mu\nu}\epsilon_{3}^{\mu\nu}\epsilon_{2\alpha\beta}\epsilon_{4}^{\alpha\beta}}{t-m^2}+\frac{ \epsilon_{1\mu\nu}\epsilon_{4}^{\mu\nu}\epsilon_{3\alpha\beta}\epsilon_{2}^{\alpha\beta}}{u-m^2}\right),
\ee
with the free coefficients in the massive gravity action chosen to be  $\kappa_3=-1$ and $\kappa_4=\frac{7}{24}$ ($c_3=\frac{1}{4}$ and $d_5=-\frac{7}{192}$ using the parametrization of \cite{deRham:2010ik}). The second term on the right hand side of \eqref{eq:M4hh} corresponds to a scalar exchange with three-point vertex given in \eqref{eq:Mphihh}. \\

Having fixed the spin-2 interactions, we then construct the scattering amplitudes for all other 2-2 scattering processes (for example $h\phi \rightarrow AA$) from the double copy prescription, and make an ansatz for the action which gives these amplitudes. A couple of general features emerge. We find that all 3 and 4 point amplitudes containing odd numbers of $A$ are zero as one would expect since $A$ is a vector. Furthermore we find that none of the amplitudes scale with energy more that $E^6$ at high energies. Since all of them have $\kappa^2=4/\mpl^{2}$ in front (can be seen from \eqref{eq:Mn}), the lowest scale appearing in the resulting theory to this order is $\Lambda_3=\left(\mpl m^2\right)^{1/3}$, the well-known highest possible scale for a Lorentz invariant theory of massive gravity. \\

As already stated, from \eqref{eq:Mhhh} and \eqref{eq:M4hh} we see that the self interactions of $h$ up to quartic order in $h$ can be described by dRGT massive gravity action. Anticipating that the $n$-point scattering amplitudes are controlled by the scale $\Lambda_3$ to all orders, it is natural to write the interactions for all the fields in the dRGT form, taking particular care to choose combinations which are natural from the point of view of the decoupling limit effective theory, namely those that automatically lead to $\Lambda_3$ interactions to all orders. This process is somewhat labourious, and we quote only our final form for the action which is
\begin{align}\label{eq:action}
\begin{split}
    S=&\int d^4x \sqrt{-g} \Bigg(\frac{2}{\kappa^2}R[g]+\frac{m^2}{\kappa^2}\sum_{n=2}^4\kappa_n\,\U_n\left[\K\right]\\
    &-\frac{1}{2}\nabla_\mu \phi \nabla^\mu \phi-\frac{1}{2}m^2 \phi^2-\frac{1}{4}F_{\mu\nu}F^{\mu\nu}-\frac{1}{2}m^2 A_{\mu}A^{\mu}\\
    -&\frac{1}{2}\K^{\mu\nu}F_{\nu\alpha}F_{\mu}^{\ \alpha}+\frac{1}{8}\K^{\mu}_{\mu}F_{\nu\alpha}F^{\nu\alpha}-\frac{1}{4}\nabla_{\mu}\phi\nabla_{\nu}\phi\left(\K^{\mu\nu}-g^{\mu\nu}\K^{\alpha}_{\alpha}\right)-\frac{\sqrt{3}}{2}\frac{m^2}{\kappa}\phi\left(\K^{\mu\nu}\K_{\mu\nu}-\K^{\mu}_{\mu}\K^{\nu}_{\nu}\right)\\
    +&\frac{1}{24\sqrt{3}}\frac{\kappa}{m^2}\phi\left([\Phi]^2-[\Phi^2]\right)+\frac{-3}{8\sqrt{3}}\kappa m^2 \phi^3-\frac{\kappa}{\sqrt{3}}m^2A^{\mu}A_{\mu}\phi-\frac{\kappa}{16\sqrt{3}}F^{\mu\nu}F_{\mu\nu}\phi+\text{quartic contact terms} \Bigg),
\end{split}
\end{align}
where $g_{\mu\nu}=\eta_{\mu\nu}+\kappa h_{\mu\nu}$ is the dynamical metric, $\eta_{\mu\nu}$ is the reference metric, $\K^{\mu}_{\nu}=\delta^{\mu}_{\nu}-(\sqrt{g^{-1}\eta })^{\mu}_{\nu}$, $\Phi_{\mu\nu}=\nabla_{\mu}\nabla_{\nu}\phi$ and the crucial contact terms which fix the form of the $2$-$2$ scattering amplitude are given in Appendix \ref{sec:contact}. The indices are raised/lowered with $g$. The self interactions of the scalar, $\phi$, contain galileon interactions (the cubic term in \eqref{eq:action} and the quartic one in \eqref{ffff}), $\phi^3$ term and two additional two and four derivative contact terms to this order. The action has been intentionally written in a manner which is diffeomorphism invariant in terms of $\K$. The reference metric $\eta$ that breaks diffeomorphism invariances only enters through $\K$, and in this sense $\K$ is a `spurion' field for the breaking of diffeomorphisms. \\

Since the $S$-matrix is invariant under field redefinitions, the cubic $\phi$ interactions are ambiguous since we may for example use field redefinitions to trade the cubic Galileon term for a potential $\phi^3$ and vice versa without changing the on-shell vertex. A similar story holds for the $\phi K^2$ and $(\nabla \phi)^2 K$ terms. However changing the off-shell structure in this way also changes the form of the quartic interactions. Anticipating that the decoupling limit is a Galileon-like theory (which is implicit in the $\Lambda_3$ scale), we have intentionally chosen to put the cubic interactions in a form for which the quartic interactions are also manifestly Galileon-like. In other words the desire to have a Galileon-like decoupling limit theory gives us guidance in writing the nonlinear off-shell structure of the theory that goes beyond what is immediately inferred from the on-shell scattering amplitudes, even though the diffeomorphism symmetry is broken by the mass term. That is the decoupling limit for the \stu fields/Goldstone modes gives us an indication of the best way to structure the interacting Lagrangian and this explains many of our choices of interactions in \eqref{eq:action} and Appendix \ref{sec:contact}. Although we have not calculated beyond four-point level, the implicit nonlinearly realized diffeomorphism symmetry present in the \stu formulation fixes a set of interactions at all orders as is familiar in effective theories with broken symmetries.

\section{$\Lambda_3$ Decoupling Limit}\label{sec:DL}

Having successfully constructed the interaction Lagrangian for the double copy effective theory, at least to quartic order, it is useful to understand its decoupling limit. This will give us insight into the interactions that arise beyond 2-2 scattering, and the overall structure of the effective theory, but it will also allow us to understand better the connection between the massive Yang-Mills decoupling limit and that for the double copy massive gravity theory. We have intentionally written the interacting Lagrangian \eqref{eq:action} in as covariant form as possible, so that the decoupling limit is easily derived. Following the standard recipe (see for example \cite{deRham:2014zqa} for a review), after denoting the reference metric from which $\K^{\mu}{}_{\nu}$ is constructed by
\be
f_{\mu\nu} = \partial_{\mu} \Phi^A \partial_{\nu} \Phi^B \eta_{AB} \, ,
\ee
we further decompose 
\be
\Phi^A = x^A - \frac{1}{m \mpl}V^A - \frac{1}{\Lambda_3^3} \eta^{AB}\partial_B \pi \, .
\ee
so that we may identify $V^A$ as the helicity-1 and $\pi$ as the helicity-0 modes of the spin-2 particle. Further for the massive spin-1 state $A_{\mu}$ we replace it by
\be
A_{\mu} \rightarrow A_{\mu} + \frac{1}{m} \partial_{\mu} \chi \, ,
\ee
where $\chi$ is the original \stu scalar, the helicity-0 state of the spin-1. The normalizations, which are standard, are chosen so that all the additional \stu fields have a finite (and non-zero) kinetic term in the decoupling limit. The metric may be denoted $g_{\mu\nu} = \eta_{\mu\nu} + \kappa h_{\mu\nu}$. Remembering that $\kappa=2/\mpl$, the decoupling limit is defined by $m \rightarrow 0$, $\kappa \rightarrow 0$ in such a way that $\Lambda_3^3 = m^2 \mpl$ is kept finite. The Lagrangian has been written in a judicious way to ensure that no term diverges in this limit. \\

Crucially, we have
\be
\lim_{m \rightarrow 0, \Lambda_3 \text{fixed}} \K_{\mu \nu} = \frac{\Pi_{\mu\nu} }{\Lambda_3^3}  := \frac{\partial_{\mu} \partial_{\nu} \pi}{\Lambda_3^3} \, ,
\ee
which explains the emergence of the Galileon symmetry for $\pi$ in the decoupling limit, since $\Pi_{\mu\nu}$ is invariant under $\pi \rightarrow \pi + c + v_{\mu}x^{\mu}$, and our choice of $\K$ as the building block. Hence for all terms in the Lagrangian for which the coefficients are finite in the $\Lambda_3$ limit, it is sufficient to replace $K_{\mu \nu} $ by $\Pi_{\mu\nu}$ and the metric $g_{\mu\nu}$ by $\eta_{\mu\nu}$. The decoupling limit Lagrangian is found to be (keeping track only of those terms which contribute to quartic order)
%DLDL
\ba \label{DLaction}
&&{\cal L}_{DL} = \frac{1}{2} h^{\mu\nu}{\cal E} h_{\mu\nu} + h_{\mu\nu} X^{\mu\nu} -\frac{1}{2}( \partial_\mu \phi)^2-\frac{1}{2} (\partial \chi)^2  +{\cal L}_{A,V}  \nn \\
   && -\frac{1}{4 \Lambda_3^3}\partial_{\mu}\phi\partial_{\nu}\phi \( \Pi^{\mu\nu}- \eta^{\mu\nu}[\Pi] \)-\frac{\sqrt{3}}{4}\frac{1}{\Lambda_3^3}\phi\left(\Pi^{\mu\nu}\Pi_{\mu\nu}-\Pi^{\mu}_{\mu}\Pi^{\nu}_{\nu}\right)  \nn  \\
    &&+\frac{1}{12\sqrt{3}}\frac{1}{\Lambda_3^3}\phi\left([\Phi]^2-[\Phi^2]\right)+\frac{11}{864}\,\frac{1}{\Lambda_3^6}\phi\left([\Phi]^3-3[\Phi][\Phi^2]+2[\Phi^3]\right)+\frac{7}{48 \Lambda_3^6}\;\varepsilon_{\mu\nu\alpha\beta}\varepsilon^{\mu'\nu'\alpha'\beta}\Phi^{\mu}_{\mu'}\;\Pi^{\nu}_{\nu'}\Pi^{\alpha}_{\alpha'}\phi \nn \\
    && +\frac{11}{8 \sqrt{3}}\,\frac{1}{\Lambda_3^6}\,\varepsilon_{\mu\nu\alpha\beta}\varepsilon^{\mu'\nu'\alpha'\beta}\Pi^{\mu}_{\mu'}\Phi^{\nu}_{\nu'}\Phi^{\alpha}_{\alpha'}\phi-\frac{11}{24\sqrt{3}}\,\frac{1}{\Lambda_3^6}\phi \left([\Pi]^3-3[\Pi][\Pi^2]+2[\Pi^3]\right) \, ,
\ea
where all indices are raised and lowered with $\eta_{\mu\nu}$.
We have separated out the spin-2 and spin-1 helicity-1 contributions which even in the case of standard massive gravity is particularly complicated \cite{Ondo:2013wka}, and they are schematically
\be
{\cal L}_{A,V}= -\frac{1}{4}F_{\mu\nu}\K^{\mu\nu\alpha \beta}F^{\alpha \beta}-\frac{1}{4}{\cal F}_{\mu\nu}{\cal K}^{\mu\nu\alpha \beta}{\cal F}^{\alpha \beta}
\ee
where ${\cal F}_{\mu \nu} = \partial_{\mu} V_{\nu}- \partial_{\nu} V_{\mu}$, $F_{\mu \nu} = \partial_{\mu} A_{\nu}- \partial_{\nu} A_{\mu}$, and the kinetic term coefficients $\K^{\mu\nu\alpha \beta}$ and ${\cal K}^{\mu\nu\alpha \beta}$ are tensors constructed from $\Pi_{\mu\nu}/\Lambda_3^3$ and $\Phi_{\mu\nu}/\Lambda_3^3$.
Since $V_{\mu}$ and $A_{\mu}$ are not sourced, classically it is consistent to set them to zero. They would of course contribute in loop processes. \\

The tensor $X_{\mu\nu}$, which is characteristic of the massive gravity decoupling limit, needs to be identically conserved to ensure that that $h_{\mu\nu}$ preserves spin-2 gauge invariance (linear diffeomorphisms) $h_{\mu\nu} \rightarrow h_{\mu\nu}+\partial_{\mu} \xi_{\nu}+\partial_{\nu} \xi_{\mu}$. This is the decoupling limit remnant of full diffeomorphism invariance. Explicitly its form is
\ba\label{Xtensor}
&&X^{aA} = \varepsilon^{abcd} \varepsilon_{ABCD} \left[  \frac{1}{2} \delta_b^B \delta_c^C  \Pi_d^D-\frac{1}{4 \Lambda_3^3} \delta_b^B   \Pi_c^C  \Pi_d^D +\frac{1}{24\Lambda_3^6}  \Pi_b^B  \Pi_c^C  \Pi_d^D + \frac{1}{24 \Lambda_3^6} \Phi_b^B  \Phi_c^C  \Pi_d^D    \right. \nn \\
&& \left. - \frac{1}{72 \sqrt{3} \Lambda_3^6} \Phi_b^B  \Phi_c^C \Phi_d^D -\frac{1}{8 \sqrt{3} \Lambda_3^6} \Phi_b^B  \Pi_c^C \Pi_d^D    \right]
\ea

The tensor \eqref{Xtensor} is indeed identically conserved by virtue of the double $\varepsilon$ structure. The full decoupling limit action \eqref{DLaction} is invariant under two separate Galileon symmetries $\pi \rightarrow \pi + v_{\mu} x^{\mu}$, $\phi \rightarrow \phi+ u_{\mu} x^{\mu}$ and thus describes a bi-Galileon theory \cite{Padilla:2010de} coupled to a massless spin-2 field. Indeed it may be put in a more manifest bi-Galileon form by performing a `demixing' transformation that removes the mixed $h \pi$ and $h \pi \pi $ terms, namely
\be
h_{\mu\nu} = \tilde h_{\mu \nu} + \frac{1}{2} \pi \delta_{\mu \nu} -\frac{1}{4 \Lambda_3^3} \pi  \Pi_{\mu\nu} \, .
\ee
We may make use of the fact that up to total derivatives
\be
 \frac{1}{2} h^{\mu\nu}{\cal E} h_{\mu\nu}  =-\frac{1}{2}  \varepsilon^{abcd} \varepsilon_{ABCD} \delta_a^A h_b^B \partial_c \partial^C h_d^D
\ee
The resulting Lagrangian then takes the form
\be
{\cal L}_{DL} = \frac{1}{2} \tilde  h^{\mu\nu}{\cal E} \tilde h_{\mu\nu} -\frac{3}{4} ( \partial \pi)^2  -\frac{1}{2} ( \partial \phi)^2-\frac{1}{2} ( \partial \chi)^2+{\cal L}^{\rm int}_{\text{bi-Galileon}}(\phi, \pi) +  \frac{1}{24 \Lambda_3^6}\varepsilon^{abcd} \varepsilon_{ABCD}   \tilde h_a^A \tilde \Pi_b^B  \tilde \Pi_c^C \tilde \Pi_d^D +{\cal L}_{A,V} \, ,
\ee
where $\tilde \Pi_{ab} = \partial_a \partial_b \tilde \pi$ and $\tilde \pi = \pi - \frac{1}{\sqrt{3}} \phi$. The term ${\cal L}^{\rm int}_{\text{bi-Galileon}}$ contains standard cubic and quartic\footnote{Strictly speaking there are also quintic interactions, however since we have only fixed the Lagrangian by reproducing the $2-2$ scattering amplitude, we cannot take seriously the inferred coefficients of the quintic interactions.} bi-Galileon interactions:
\be
{\cal L}^{\rm int}_{\text{bi-Galileon}} =  a_0 \pi ( \varepsilon \varepsilon \delta^2 \Pi^2) + \phi \sum_{n=1}^3 a_n (  \varepsilon \varepsilon  \delta^2 \Phi^{n-1} \Pi^{3-n}  ) + b_0 \pi ( \varepsilon \varepsilon \delta \Pi^3)  + \phi \sum_{n=1}^4 b_n (  \varepsilon \varepsilon  \delta \Phi^{n-1} \Pi^{4-n}  ) \, , 
\ee
where we have used the shorthand $\epsilon \epsilon XYZW = \varepsilon^{abcd} \varepsilon_{ABCD} X_a^A Y_b^B Z_c^C W_d^D$ and the coefficients are given by $(a_0,a_1,a_2,a_3)= (-\frac{1}{8},\frac{\sqrt{3}}{8},-\frac{1}{8},\frac{1}{24 \sqrt{3}})$ and $(b_0,b_1,b_2,b_3,b_4)=(\frac{5}{96},-\frac{25 \sqrt{3} }{144} , \frac{1}{6}, \frac{197 \sqrt{3}}{432} ,  \frac{11}{864})$. \\

The quartic interactions of the form $\tilde h \tilde \Pi^3$ cannot be removed with a local field redefinition, as is well known from the standard massive gravity case. This is as it should be since it is precisely these interactions that describe the nonzero helicity $0,0,0,\pm 2$ amplitudes that arise from the $\Sigma \Sigma'$ contact term in the decoupling limit, as described in equation \eqref{eq:gravityDL} and implicit in the full answer \eqref{eq:fullamp} and explicit in \eqref{eq:tensoramp}. Indeed the combination $\tilde \pi$ is exactly the combination which identifies the diagoanlized parts of $\pi$ and $\phi$ that correspond to the spin-1 helicity-0 polarization tensor squared $\epsilon_0^{\mu} \epsilon_0^{\nu}$\footnote{To see this, note that at leading order in the decoupling limit $K_{\mu\nu} \sim \frac{1}{\Lambda_3^3}\partial_{\mu} \partial_{\nu} \pi \sim - \sqrt{3/2} \frac{1}{\mpl}\epsilon_{\mu\nu}^{\lambda=0} \pi$. Since in unitary gauge $K_{\mu\nu}=\frac{1}{\mpl}h_{\mu\nu}+{\cal O} (h^2)$, the canonically normalizalized unitary gauge helicity-zero mode is in effect $- \sqrt{3/2} \pi$, whence the combination arising in \eqref{eq:comb} is $\frac{2}{\sqrt{6}} ( - \sqrt{3/2} \pi+ \frac{1}{\sqrt{2}}\phi)=- \pi + \frac{1}{\sqrt{3} \phi}= - \tilde \pi$.  }. \\

As noted in the introduction, since the decoupling limit of massive Yang-Mills is a nonlinear sigma model and the double copy of the latter is the special Galileon, we might have expected the massive gravity theory to be that corresponding to a special Galileon. Interestingly however, this was never possible since the decoupling limit of dRGT massive gravity never gives rise to a special Galileon. This is easily seen by the manner in which the Galileon interactions arise from mixing with $h_{\mu\nu}$. The decoupling limit of dRGT massive gravity for general $\kappa_3$ and $\kappa_4$ is (ignoring helicity-1 contributions)
\be
{\cal L}_{DL} =- \frac{1}{2}  \varepsilon^{abcd} \varepsilon_{ABCD} \delta_a^A h_b^B \partial_c \partial^C h_d^D+ h_{\mu\nu} X^{\mu\nu} \, ,
\ee
where
\be
X_{\mu\nu} =  \varepsilon^{abcd} \varepsilon_{ABCD} \left[  \frac{1}{2} \delta_b^B \delta_c^C  \Pi_d^D + \frac{1}{4 \Lambda_3^3}(2+3 \kappa_3) \delta_b^B   \Pi_c^C  \Pi_d^D +\frac{1}{4\Lambda_3^6} (4 \kappa_4+\kappa_3)  \Pi_b^B  \Pi_c^C  \Pi_d^D \right] \, .
\ee
Since the special Galileon in four dimensions is a pure quartic Galileon, we need that after performing the demixing 
\be
h_{\mu\nu}= \tilde h_{\mu\nu} + \frac{1}{2} \pi \eta_{\mu\nu} + \frac{1}{4 \Lambda_3^3}(2+3 \kappa_3) \pi \Pi_{\mu\nu} \, ,
\ee
there is no cubic Galileon term. This requires $(2+3 \kappa_3)=0$ which does not correspond to the value obtained from double copy. Even with this choice, we then have
\ba
{\cal L}_{DL} &=&- \frac{1}{2}  \varepsilon^{abcd} \varepsilon_{ABCD} \delta_a^A \tilde h_b^B \partial_c \partial^C \tilde h_d^D+ \frac{1}{8\Lambda_3^6} \varepsilon^{abcd} \varepsilon_{ABCD}  (4 \kappa_4+\kappa_3)   \pi  \delta_a^A \Pi_b^B  \Pi_c^C  \Pi_d^D \nn  \\
&&+\frac{1}{4\Lambda_3^6} (4 \kappa_4+\kappa_3) \varepsilon^{abcd} \varepsilon_{ABCD}  \tilde h_a^A \Pi_b^B  \Pi_c^C  \Pi_d^D  \, ,
\ea
and so we only have a non-vanishing quartic Galileon term when there is also a non-zero $h\pi\pi \pi$ interaction which cannot itself be removed with a field redefinition since it contributes to the $\pm2,0,0,0$ scattering amplitude. Furthermore higher order $n$-point amplitudes will receive contributions from intermediate graviton exchange which do not arise in the pure quartic Galileon theory. Hence the special Galileon does not strictly speaking arise in standard massive gravity in any form.

\section{Discussion}
In this paper we explored the possibility of constructing an interacting massive spin-2 theory, {\it i.e.} massive gravity, as a double copy of massive Yang-Mills theory. 
Our prescription for doing this is to demand that the kinematic factors for the massive theory, defined by normalizing by the massive scalar propagator \eqref{eq:An}, satisfy the same colour/kinematics duality as the massless case. This is a nontrivial requirement even at the level of 2-2 scattering, we nevertheless find that it remains intact to this order. Interestingly the ambiguity that arises in the massless case (e.g. the ability to shift $n_s$ to $n_s+\alpha s$ {\it etc.} - the so-called generalized gauge transformations) is fixed in the massive case by the requirement that colour kinematics holds. Furthermore the manner in which it is fixed is such that the kinematic factors contain a term which is singular in the decoupling limit, but nevertheless leads to finite contributions to the gravity amplitudes. One consequence of this is that the decoupling limit and double copy procedures do not commute, a result which could not have been anticipated from the decoupling limit theories alone. Hence the by now well known relations between the scattering amplitudes of nonlinear sigma models, special Galileons etc \cite{Cachazo:2014xea,Cheung:2017ems,Cheung:2017yef, Cheung:2016prv} which appear to be part of a large web of interconnected theories, are non-trivially lifted by the presence of a mass term. \\

It is beyond the scope of this paper to consider higher $n$-point amplitudes which are needed to check whether the double copy procedure remains intact at all orders, however it was shown in \cite{Johnson:2020pny} that at 5 points the double copy of massive Yang-Mills amplitude gives spurious poles and therefore that cannot be matched with an amplitude calculated from a local Lagrangian. The reason for such poles is that at 5 point the massive Yang-Mills kinematic factors, $n_i$, calculated directly from Feynman rules do not satisfy Jacobi identities, so they need to be shifted as $n_i\rightarrow n_i+\Delta_i$ so that the amplitude remains unchanged. As it turns out for generic theories such shifts are non local, i.e. they have poles in kinematic invariants, $s_{ij}$. For, example in massive Yang-Mills case these shifts contain the following polynomial of $s_{ij}$ in the denominator
\begin{align}
&\,320 m^8+36 m^6 (9 s_{12}+4 (s_{13}+s_{14}+s_{23}+s_{24}))\nonumber\\
&\hspace{-0.5cm}+m^4 \left(117 s_{12}^2+108 s_{12} (s_{13}+s_{14}+s_{23}+s_{24})+4 \left(s_{13} (13 s_{14}+4 s_{23}+17 s_{24})\right.\right.\nonumber\\
&\hspace{1cm}\left.\left.+4
s_{13}^2+4 s_{14}^2+17 s_{14} s_{23}+4 s_{14} s_{24}+4 s_{23}^2+13 s_{23} s_{24}+4
s_{24}^2\right)\right)\nonumber\\
&\hspace{-0.5cm}+2 m^2 \left(9 s_{12}^3+13 s_{12}^2 (s_{13}+s_{14}+s_{23}+s_{24})+s_{12} \left(s_{13} (10 s_{14}+6 s_{23}+17
s_{24})\right.\right.\nonumber\\
&\hspace{1cm}\left.\left.+4 s_{13}^2+4 s_{14}^2+s_{14} (17 s_{23}+6 s_{24})+2 (2 s_{23}+s_{24}) (s_{23}+2 s_{24})\right)\right.\nonumber\\
&\hspace{1cm}\left.+2 \left(s_{13}^2 (s_{14}+2 s_{24})+s_{13}
\left(s_{14}^2+s_{14} (s_{23}+s_{24})+s_{24} (s_{23}+2 s_{24})\right)\right.\right.\nonumber\\
&\hspace{1cm}\left.\left.+s_{23} \left(s_{24} (s_{14}+s_{23})+2 s_{14}
(s_{14}+s_{23})+s_{24}^2\right)\right)\right)\nonumber\\
&\hspace{-0.5cm}+2 s_{24} \left(s_{23} \left(s_{12}^2+s_{12} (s_{13}+s_{14})-s_{13} s_{14}\right)+s_{12}
(s_{12}+s_{13}) (s_{12}+s_{13}+s_{14})\right)\nonumber\\
&\hspace{-0.5cm}+(s_{12} (s_{12}+s_{13}+s_{14})+s_{23} (s_{12}+s_{14}))^2+s_{24}^2
(s_{12}+s_{13})^2,
\end{align}
which has some complicated expressions of zeros which are the locations of unphysical poles when the shifted $n_i$'s are squared. In \cite{Johnson:2020pny} the conditions on the spectrum of the theory for avoiding such poles are derived and it was shown that massive Yang-Mills theory does not satisfy them. Therefore, the double copy of massive Yang-Mills amplitudes can only be matched with amplitudes calculated from the gravitational action in \eqref{eq:action} only at three and four points. While this result is clearly negative in terms of the naive application of the conjecture \eqref{eq:Mn}, we do not regard this as necessarily terminal for several reasons. It is worth noting that in the present context there is more freedom that the conventional story because both sides of the double copy are effective theories. We are free to add irrelevant operators suppressed by the scale $\Lambda$ or a parametrically lower scale on the Yang-Mills side, for the sole purpose of ensuring the colour/kinematics relation remains intact even when the spectral conditions of \cite{Johnson:2020pny} are not satisfied. One reasonable conjecture is that this should be possible in such a manner to ensure that the gravitational theory remains a $\Lambda_3$,  or similarly at a parameterically lower scale, theory to all orders. Some support for this comes from the fact that if we only focus on those amplitudes that arise from helicity-0 modes of the spin-1, then the double copy procedure is known to work to all orders since it gives a special Galileon for which all interactions arise at the scale $\Lambda_3$. Since as we have seen the decoupling limit and double copy procedures do not commute, this does not constitute a proof. \\

Perhaps more importantly though, the rules of the double copy paradigm for massive theories, even if they apply, are not well established and it may not be meaningful to impose the conjecture \eqref{eq:Mn} strictly to all orders. Since nearly all cases in which the double copy has been well established are for massless theories, it is not unreasonable to suppose that something like equation \eqref{eq:Mn} may only be true at leading order in an expansion in powers of $m$, i.e. at leading order in the decoupling limit. What we have been able to construct, as outlined in section~\ref{sec:DL}, is a local $\Lambda_3$ effective theory whose interactions match with the double copy prescription with massive Yang-Mills up to 4-point order. The effective theory whose leading terms are given in \eqref{eq:action} is consistent to all orders (from the low energy point of view) and we may use it to compute arbitrary local (analytic) $n$-point functions at a given order in an EFT expansion, even if these $n$-point functions do not precisely match those implied by the double copy conjecture \eqref{eq:Mn}. That is we can infer the leading contributions to the higher point interactions in \eqref{eq:action} from the decoupling limit rather than double copy. That this is possible despite having only computed up to 4-point order is due to the role played by the nonlinearly realized symmetry, from the underlying symmetry breaking in giving mass, and how it determines the leading interactions of the low energy theory, organizing the structure of the EFT. It remains possible that the effective theory outlined in \eqref{eq:action}, or some close relative to it, does have a role to play in an appropriately double copied spontaneously broken gauge theory. \\

It is also worth remembering that the de Rham-Gabadadze-Tolley model of massive gravity is really just one example of large family of interacting effective theories which may contain any number of massive spin-2 and lower states just as massive Yang-Mills is just one example of an effective theory for a spontaneously broken gauge symmetry. It would be interesting to explore extensions on both sides to understand to what extent the double copy paradigm may be preserved either fully in the sense of a strict relation along the lines of \eqref{eq:Mn}, or at least partially in a weaker form. The present work serves as a starting point for such an analysis and for establishing the relation between the decoupling limits on each side. What is interesting to note is that the class of Galileon effective theories emerge ubiquitously in decoupling limits of these theories. Indeed the original Galileon model was first noted to arise in a diffeomorphism invariant five dimensional theory, the Dvali-Gabadadze-Porrati model \cite{Luty:2003vm} which is superficially quite different to ghost-free massive gravity. There the effective four dimensional graviton emerges as a resonance, i.e. may be viewed as a continuum of massive states. The intimate connection between the NLSM and the special Galileon, and the similar emergence of the bi-Galileon effective theory in our present discussion suggest that there may be an analogous double copy prescription to \eqref{eq:Mn} which could be applied to soft massive theory in which the mass for the spin-1 field on the gauge side emerges from a resonance. In this manner the spurious pole issue identified for higher $n$-point functions in \cite{Johnson:2020pny} may be resolved by satisfying constraints on the interactions of the continuum modes without needing to satisfy specific spectrum conditions since the spectrum itself is continuous.
\\

There are many interesting future directions that these results suggest. For example, is it possible to add irrelevant operators to \eqref{eq:lagr mYM} or new fields such that the double copy procedure works for all $n$-point amplitudes, or is it only possible if the spectral conditions of \cite{Johnson:2020pny} are satisfied?  If no extra fields or higher order operators can fix the spurious poles then maybe this suggests that there is some problem of constructing such a massive gravitational theory elsewhere. For example, as an anonymous referee pointed out, for Yang-Mills theory with a single adjoint fermion the colour kinematics duality cannot be satisfied in more than 10 dimensions because gravitational theory with more than 8 gravitini and no supersymmetry cannot exist in 4 dimensions. Also, can we construct the double copy of Yang-Mills action coupled to a Higgs field, which is the UV completion of massive Yang-Mills considered here, and what is the resulting gravitational theory? \footnote{The double copy of two spontaneously broken Yang-Mills theories where all fields have mass $m$ was studied in \cite{Chiodaroli:2015rdg} and spurious poles $s-(2m)^2$ were found even in 4pt amplitudes. It was suggested these poles show the existence of a state of mass $2m$ which has to be added to the spectrum. Therefore, an interesting future direction would be to understand if the 5pt spurious poles in massive Yang-Mills signal a presence of another state as well.}  Does this give us any insight into UV completing massive gravity theories? Does the procedure outlined hold at loop level in any way? Are there some simple extensions of the classical double copy relations? The latter would be highly nontrivial given the known complicated nonlinear dynamics of massive gravity theories exhibited by the Vainshtein mechanism. We leave these various considerations to future work.

\bigskip
\noindent{\textbf{Acknowledgments:}}
We would like to thank Mariana Carrillo Gonzalez, Clifford Cheung, Claudia de Rham, Mike Duff and Laura Johnson for useful comments. The work of AJT is supported by an STFC grant ST/P000762/1. JR is supported by an STFC studentship. AJT thanks the Royal Society for support at ICL through a Wolfson Research Merit Award.

\newpage
\appendix
\section{Contact terms}\label{sec:contact}

%\subsection{$\L^{(4)}_{A\phi A\phi}$}
Below are the various contact terms needed in \eqref{eq:action} to reproduce the desired quartic interactions. All terms are written in a covariant form, with the understanding that they enter the action with a $\sqrt{-g}$ prefactor.

\begin{align}\label{ffff}
\begin{split}
    \L^{(4)}_{\phi\phi\phi\phi}= &\frac{11}{3456}\,\frac{\kappa^2}{m^4}\phi\left([\Phi]^3-3[\Phi][\Phi^2]+2[\Phi^3]\right)+\frac{21}{128}\,\kappa^2\nabla^{\mu}\phi\nabla_{\mu}\phi\;\phi\;\phi+\frac{-1}{96}\,\frac{\kappa^2}{m^2}\nabla^{\rho}\phi\nabla^{\sigma}\phi\Phi_{\rho\sigma}\;\phi
\end{split}    
\end{align}
where we have used $\Phi_{\mu\nu}=\nabla_{\mu}\nabla_{\nu}\phi$.
\begin{align}
\begin{split}
    \L^{(4)}_{AAAA}= &\frac{-1}{512}\,\frac{\kappa^2}{m^2}(F^{\mu\nu}F_{\mu\nu})^2+\frac{3}{256}\,\frac{\kappa^2}{m^2}F^{\mu\nu}F_{\nu\rho}F^{\rho\sigma}F_{\sigma\mu}
\end{split}    
\end{align}
%\subsection{$\L^{(4)}_{AAhh}$}

\begin{align}
\begin{split}
  \L^{(4)}_{AAhh}= &\frac{-3}{16}\;F_{\mu\nu}F_{\rho\sigma}\K^{\mu\rho}\K^{\nu\sigma}+\frac{-1}{4}\;F^{\mu\nu}F_{\mu\sigma}\K_{\nu\rho}\K^{\rho\sigma}+\frac{-1}{16}\;\varepsilon_{\mu\nu\alpha\beta}\varepsilon^{\mu'\nu'\alpha'\beta'}F^{\mu}_{\mu'}F^{\nu}_{\nu'}\K^{\alpha}_{\alpha'}\K^{\beta}_{\beta'}\\
    \end{split}    
\end{align}

%\subsection{$\L^{(4)}_{hh\phi\phi}$}

\begin{align}
\begin{split}
   \L^{(4)}_{hh\phi\phi}= &\frac{7}{48}\;\varepsilon_{\mu\nu\alpha\beta}\varepsilon^{\mu'\nu'\alpha'\beta}\Phi^{\mu}_{\mu'}\;\K^{\nu}_{\nu'}\K^{\alpha}_{\alpha'}\phi+\frac{3}{8}\;\varepsilon_{\mu\nu\alpha\beta}\varepsilon^{\mu'\nu'\alpha'\beta}\nabla^{\mu}\phi\nabla_{\mu'}\K^{\nu}_{\nu'}\K^{\alpha}_{\alpha'}\phi\\

+&\frac{-17}{48}\;m^2\;\phi\;\phi\;\K^{\mu\nu}\K_{\mu\nu}+\frac {1} {24}\;\frac{1}{m^2}\;\varepsilon_{\mu\nu\alpha\beta}\varepsilon^{\mu'\nu'\alpha'\beta'}\nabla^{\mu}\phi\;\nabla_{\mu'}\phi\;\nabla^{\nu}\K^{\alpha}_{\alpha'}\nabla_{\nu'}\K^{\beta}_{\beta'}\\

+&\frac{1}{12}\;\frac{1}{m^2}\;\varepsilon_{\mu\nu\alpha\beta}\varepsilon^{\mu'\mu\alpha'\beta'}\nabla^{\rho}\phi\;\nabla^{\nu}\phi\;\nabla_{\mu'}\K^{\alpha}_{\alpha'}\nabla_{[\rho}\K^{\beta}_{\beta']}+\frac{1}{48}\;\frac{1}{m^2}\;\varepsilon_{\mu\nu\alpha\beta}\varepsilon^{\mu'\nu'\alpha'\beta'}\Phi^{\mu}_{\mu'}\Phi^{\nu}_{\nu'}\K^{\alpha}_{\alpha'}\K^{\beta}_{\beta'}
\end{split}
\end{align}

%\subsection{$\L^{(4)}_{h\phi\phi\phi}$}

\begin{align}
\begin{split}
    \L^{(4)}_{h\phi\phi\phi}=&\frac{-1}{144\sqrt{3}}\,\frac{\kappa}{m^4}\,\varepsilon_{\mu\nu\alpha\beta}\varepsilon^{\mu'\nu'\alpha'\beta'}\K^{\mu}_{\mu'}\Phi^{\nu}_{\nu'}\Phi^{\alpha}_{\alpha'}\Phi^{\beta}_{\beta'}+\frac{-19}{48\sqrt{3}}\,\kappa\, \K^{\mu\nu} \nabla_{\mu}\phi\nabla_{\nu}\phi\;\phi\\
+&\frac{11}{16\sqrt{3}}\,\frac{\kappa}{m^2}\,\varepsilon_{\mu\nu\alpha\beta}\varepsilon^{\mu'\nu'\alpha'\beta}\K^{\mu}_{\mu'}\Phi^{\nu}_{\nu'}\Phi^{\alpha}_{\alpha'}\phi
   \end{split}
\end{align}

%\subsection{$\L^{(4)}_{hhh\phi}$}

\begin{align}
\begin{split}
     \L^{(4)}_{hhh\phi}=&\frac{1}{12\sqrt{3}}\,\frac{1}{\kappa}\,\varepsilon_{\mu\nu\alpha\beta}\varepsilon^{\mu'\nu'\alpha'\beta'}\K^{\mu}_{\mu'}\K^{\nu}_{\nu'}\K^{\alpha}_{\alpha'}\Phi^{\beta}_{\beta'}+\frac{-2}{\sqrt{3}}\,\frac{1}{\kappa}\,\nabla^{[\beta}\K^{\nu]\alpha}\nabla_{[\beta}\K_{\mu]\alpha}\K^{\mu}_{\nu}\phi\\

+& \frac{8}{ \sqrt{3}}  \frac{1}{\kappa} R_{\mu\nu}^{\rho\sigma}\K^{\mu}_{\rho}\K^{\nu}_{\sigma}\phi+\frac{-11}{12\sqrt{3}}\,\frac{m^2}{\kappa}\phi\left([\K]^3-3[\K][\K^2]+2[\K^3]\right)\end{split}    
\end{align}
with $\nabla_{[\mu}A_{\nu]\rho}=\frac{1}{2}(\nabla_{\mu}A_{\nu\rho}-\nabla_{\nu}A_{\mu\rho})$ 

\begin{align}
\begin{split}
     \L^{(4)}_{AhA\phi}=&\frac{1}{8\sqrt{3}}\,\kappa m^2\,A^{\mu}A^{\nu}\K_{\mu\nu}\phi+\frac{1}{16\sqrt{3}}\,\frac{\kappa}{m^2}\,F^{\mu\nu}F^{\rho\sigma}\K_{\nu\rho}\Phi_{\mu\sigma}+\frac{-1}{16\sqrt{3}}\,\frac{\kappa}{m^2}\,\nabla^{\rho}F^{\mu\nu}\nabla_{\sigma}F_{\mu\nu}\K_{\nu}^{\sigma}\phi\\+
&\frac{-1}{4\sqrt{3}}\,\frac{\kappa}{m^2}\,\nabla^{\nu}F^{\mu\rho}F_{\mu\sigma}\nabla_{\rho}\K_{\nu}^{\sigma}\phi+\frac{-1}{8\sqrt{3}}\,\frac{\kappa}{m^2}\,\nabla^{\rho}F^{\nu}_{\sigma}\nabla_{\rho}F^{\mu\sigma}\K_{\mu\nu}\phi+\frac{-1}{8\sqrt{3}}\,\frac{\kappa }{m^2}\,F_{\mu\nu}F_{\rho\sigma}\nabla^{\mu}\nabla^{\sigma}\K^{\nu\rho}\phi\\
\end{split}    
\end{align}
%\subsection{$\L^{(4)}_{A\phi A\phi}$}

\begin{align}
\begin{split}
    \L^{(4)}_{A\phi A\phi}= &\frac{1}{384}\,\frac{\kappa^2}{m^4}\,F^{\mu\nu}F_{\mu\nu}\Phi^{\rho\sigma}\Phi_{\rho\sigma}+\frac{1}{64}\,\kappa^2\,F^{\mu\nu}F_{\mu\rho}\phi\phi+\frac{-1}{48}\,\frac{\kappa^2}{m^2}\,F^{\mu\nu}F_{\nu}^{\rho}\nabla_{\rho}\phi\nabla_{\mu}\phi\\
     +&\frac{-11}{32}\,\kappa^2m^2\,A^{\mu}A_{\mu}\phi\phi+\frac{1}{192}\,\kappa^2\,A^{\mu}A^{\nu}\nabla_{\mu}\phi\nabla_{\nu}\phi\\
     +&\frac{-1}{192}\,\frac{\kappa^2}{m^4}\,\nabla^{\rho}F_{\mu\nu}\nabla^{\sigma}F^{\mu\nu}\nabla_{\rho}\phi\nabla_{\sigma}\phi+{\frac{1}{128}}\,\frac{\kappa^2}{m^2}\,\nabla_{\rho}F_{\mu\nu}\nabla^{\rho}F^{\mu\nu}\phi\phi\\
     +&\frac{-1}{192}\,\frac{\kappa^2}{m^4}\,\nabla^{\rho}F^{\mu}_{\nu}\nabla_{\mu}F_{\rho}^{\sigma}\nabla^{\nu}\phi\nabla_{\sigma}\phi+\frac{1}{96}\,\frac{\kappa^2}{m^4}\,\nabla^{\rho}F^{\mu\nu}\nabla_{\sigma}F_{\nu}^{\sigma}\nabla_{\rho}\phi\nabla_{\mu}\phi
\end{split}    
\end{align}

\section{Conventions}\label{sec:conv}
\subsection{Lie Algebra Generators of the Gauge Group}
We use the following conventions for the generators, $T_a$:
\be
Tr[T_aT_b]=\delta_{ab}.
\ee
These are related to the usual generators (for example in \cite{peskin}) as $T_a=\sqrt{2}t_a$. We define the structure constants, $f_{abc}$ as
\be
[T_a,T_b]=i f_{abc} T_c,
\ee
which again are larger by a factor of $\sqrt{2}$ than the structure constants in \cite{peskin}. In terms of these $f_{abc}$ the field strength tensor $F^a_{\mu\nu}$ is written as:
\be
F^a_{\mu\nu}=\partial_{\mu}A^a_{\nu}-\partial_{\nu}A^a_{\mu}+\frac{1}{\sqrt{2}}f_{abc}A^{b}_{\mu}A^{c}_{\nu}.
\ee

\subsection{Polarizations}
The four momenta in the centre of mass frame with scattering angle $\theta$ and three momenta $p=\frac{1}{2}\sqrt{s-4m^2}$ is defined as:
\begin{equation}
    p^{\mu}=(\frac{\sqrt{s}}{2}, p\sin{\theta}, 0, p\cos{\theta}).
\end{equation}
We define the polarization vectors in the helicity basis as follows:
\begin{align}\label{eq: pol vectors}
\begin{split}
   &\epsilon^{\mu}_{\lambda=1} = \frac{1}{\sqrt{2}}(0,-\cos{\theta},-i,\sin{\theta})\,, \\
   &\epsilon^{\mu}_{\lambda=-1} = \frac{1}{\sqrt{2}}(0,\cos{\theta},-i,-\sin{\theta})\,, \\
    &\epsilon^{\mu}_{\lambda=0} = \frac{1}{m}(p,E\sin{\theta},0,E\cos{\theta})\,.
\end{split}
\end{align}
where $\theta$ is the scattering angle in the centre of mass frame and $p$ the three-momentum. The polarizations clearly satisfy the transverse and completeness relations,$i.e$

\begin{align}\label{eq:pol req}
\begin{split}
    p_{\mu}\epsilon^{\mu}_{\lambda}&=0\;,\\
    \sum_{\lambda=1}^{3}\epsilon^{\mu}_{\lambda}(\epsilon^{\nu}_{\lambda})^*&=\eta^{\mu\nu}+\frac{p^\mu p^\nu}{m^2}\;,
\end{split}
\end{align}
where $(\epsilon^{\mu}_{\lambda})^*=(-1)^{\lambda}\epsilon^{\mu}_{-\lambda}$. The polarization tensors for the spin-2 field with different helicities are constructed from the polarization vectors  with  appropriate Clebsch-Gordan (CG) coefficients as (we review the construction in detail in \ref{sec:CG}),
\begin{align}
\begin{split}
    &\epsilon^{\mu\nu}_{\lambda=\pm 2} = \epsilon^{\mu}_\pm    \epsilon^{\nu}_\pm\,, \\
    &\epsilon^{\mu\nu}_{\lambda=\pm 1} = \frac{1}{\sqrt{2}}(\epsilon^{\mu}_\pm\epsilon^{\nu}_0 +
   \epsilon^{\mu}_0\epsilon^{\nu}_\pm)\,, \\
  &\epsilon^{\mu\nu}_{\lambda=0} = \frac{1}{\sqrt{6}}
    (\epsilon^{\mu}_+\epsilon^{\nu}_-
    + \epsilon^{\mu}_-\epsilon^{\nu}_+
    + 2\epsilon^{\mu}_0\epsilon^{\nu}_0)\,.
\end{split}
\end{align}
The polarization tensors satisfy the transverse, traceless and completeness relations

\begin{align}
\begin{split}
    &p_{\mu}\epsilon^{\mu\nu}_{\lambda}=0\;,\epsilon^{\mu}_{\mu\lambda}=0\;,\\
    &\sum_{\lambda=-2}^{2}\epsilon^{\mu\nu}_{\lambda}(\epsilon^{\alpha\beta}_{\lambda})^{*}=\frac{1}{2}\left(G^{\mu\alpha}G^{\nu\beta}+G^{\mu\beta}G^{\nu\alpha}-\frac{2}{3}G^{\mu\nu}G^{\alpha\beta}\right),
\end{split}
\end{align}
where $G^{\mu\alpha}=\eta^{\mu\nu}+\frac{p^\mu p^\nu}{m^2}$.

\subsection{Construction of gravity states from mYM}\label{sec:CG}

As mentioned in \ref{sec:dof}, from the tensor product of two massive spin-1 states we get a massive spin-2, a massive spin-1 and a massive spin-0 on the gravity side. In this section we review how the gravity on-shell states are constructed from such product, $i.e$, $|1,\lambda_1>\otimes|1,\lambda_2>$. The polarization tensor of the particle of spin $J$ with helicity $\lambda$ is given as,

\begin{equation}
\epsilon^{J,\lambda}_{\mu\nu}=\sum_{\lambda'\lambda''}C_{\lambda'\lambda''}^{J,\lambda}\epsilon^{\lambda'}_{\mu}\epsilon^{\lambda''}_{\nu},
\end{equation}
where $\lambda=\lambda'+\lambda''$. We start from the spin-0 state which is obtained from $|0,\lambda>=|1,\lambda'>\otimes|1,\lambda''>$, with $\lambda=0=\lambda'+\lambda''$. This polarization state is obtained by considering the following:

\begin{align}\label{pol f}
    \begin{split}
     \epsilon^{(\phi)}_{\mu\nu}\equiv\epsilon^{0,0}_{\mu\nu} =&\sum_{\lambda'\lambda''}C_{\lambda'\lambda''}^{0,0}\epsilon^{\lambda'}_{\mu}\epsilon^{\lambda''}_{\nu}\\
        =&\frac{1}{\sqrt{3}}\left(\epsilon^{0}_{\mu}\epsilon^{0}_{\nu}-\epsilon^{+}_{\mu}\epsilon^{-}_{\nu}-\epsilon^{-}_{\mu}\epsilon^{+}_{\nu}\right)
    \end{split}
\end{align}
where $C_{\lambda'\lambda''}^{0}$ are the CG coefficients given in \eqref{eq: CG}. By substituting \eqref{eq: pol vectors}, we can see that \eqref{pol f} can be expressed as:
\begin{align}
    \begin{split}
        \epsilon^{(\phi)}_{\mu\nu}=\frac{1}{\sqrt{3}}\left(\eta_{\mu\nu}+\frac{p_\mu p_\nu}{m^2}\right).
    \end{split}
\end{align}
Hence, the factor of $\frac{1}{\sqrt{3}}$ in \eqref{eq:pol phi} which follows from the CG coefficient.\\

The spin-2 state is obtained from $|2,\lambda>=|1,\lambda'>\otimes|1,\lambda''>$, with $-2\leq\lambda\leq2$.

\begin{align}
    \begin{split}
     \epsilon^{2,\lambda}_{\mu\nu} =&\sum_{\lambda'\lambda''}C_{\lambda'\lambda''}^{2,\lambda}\epsilon^{\lambda'}_{\mu}\epsilon^{\lambda''}_{\nu}.
    \end{split}
\end{align}
 To give an explicit example, the helicity $\lambda=+2$ is,
 \begin{align}
    \begin{split}
     \epsilon^{2,+2}_{\mu\nu} =&\sum_{\lambda'\lambda''}C_{\lambda'\lambda''}^{2,+2}\epsilon^{\lambda'}_{\mu}\epsilon^{\lambda''}_{\nu},\\
     =&C_{+1+1}^{2,+2}\epsilon^{+1}_{\mu}\epsilon^{+1}_{\nu},\\
     =&1\times\epsilon^{+1}_{\mu}\epsilon^{+1}_{\nu}.
    \end{split}
\end{align}
 In this paper we use the polarization states to be a superposition of different helicities and we do not focus on specific choices, for example for the graviton polarisation we have,
 
\begin{equation}
 \epsilon^{(h)}_{\mu\nu}=\sum_{\lambda=-2}^{+2}\alpha_{\lambda}\epsilon^{2,\lambda}_{\mu\nu}.
\end{equation}

 \begin{align}\label{eq: CG}
    \begin{split}
     \text{Spin-2}:\quad&C_{++}^{2,2}=C_{--}^{2,-2}=1,\\
     &C_{0+}^{2,1}=C_{+0}^{2,1}=C_{-0}^{2,-1}=C_{0-}^{2,-1}=\frac{1}{\sqrt{2}},\\
     &C_{+-}^{2,0}=C_{-+}^{2,0}=\frac{1}{\sqrt{6}},\;\quad C_{00}^{2,0}=\sqrt{\frac{2}{3}},\\
    \text{Spin-1}:\quad &C_{+0}^{1,1}=-C_{0+}^{1,1}=C_{0-}^{1,-1}=-C_{-0}^{1,-1}=\frac{1}{\sqrt{2}}\\
     &C_{++}^{1,1}=C_{++}^{1,1}=\frac{1}{\sqrt{2}},\;\quad C_{00}^{1,0}=0\\
    \text{Spin-0}:\quad &C_{+-}^{0,0}=C_{-+}^{0,0}=\frac{-1}{\sqrt{3}},\;\quad C_{00}^{0,0}=\frac{1}{\sqrt{3}}
     
    \end{split}
\end{align}
 
\section{Dualization of the massive $B$ field in 4d}\label{sec:dual}
We follow the dualization procedure explained in \cite{ortin}. The \stu action of free massive 2-form field, $B$, is 
\be
S=\int -\frac{1}{2}dB\wedge *dB-\frac{1}{2}(m B-d \lambda)\wedge *(m B-d \lambda),
\ee
where $\lambda$ is a 1-form \stu field which is needed to restore the gauge symmetry which acts on the fields as follows:
\begin{align*}
&B\rightarrow B + d\Lambda,\\
&\lambda\rightarrow \lambda+m \Lambda. 
\end{align*}
The first step in the dualization procedure is to rewrite the action in terms of field strengths, $H=dB$ and $G=mB-d\lambda$. To do that we need to impose Bianchi identities,
\begin{align}
    &dH=0,\label{eq:dH}\\
    &dG-mdB=0,\label{eq:dG}
\end{align}
with Lagrange multipliers. We first do it for $G$:
\be
S=\int -\frac{1}{2}dB\wedge *dB-\frac{1}{2}G\wedge *G+A\wedge d(G-mB),
\ee
where $A$ is a 1-form Lagrange multiplier imposing \eqref{eq:dG}. By integrating the last term by parts we can find the equation of motion for $G$ to be
\be
G=-*dA.
\ee
Substituting this back to the action and integrating by parts the last term we get
\be
S=\int -\frac{1}{2}dB\wedge *dB-\frac{1}{2}dA\wedge *dA+A\wedge mdB.
\ee
Now we can replace $dB$ by $H$ and impose \eqref{eq:dH} with a scalar Lagrange multiplier, $\chi$. This gives the following
\be\label{eq:S_HHAAchi}
S=\int -\frac{1}{2}H\wedge *H-\frac{1}{2}dA\wedge *dA+A\wedge mH+\chi dH.
\ee
Now again we integrate last term by parts and find the equation of motion for $H$ to be 
\be\label{eq:HA}
H=-*(m A-d\phi).
\ee
Substituting this back in the \eqref{eq:S_HHAAchi} gives the Stueckelberg action for massive spin-1 field, $A$, known as Proca action:
\be
S=\int -\frac{1}{2}dA\wedge *dA-\frac{1}{2}(mdA-d\chi)\wedge *(mdA-d\chi),
\ee
where $\chi$ is now the \stu scalar field. From \eqref{eq:HA} we can see that in unitary gauge, $\chi=0$, the relation between the $B$ and $A$ fields is $dB=-*mA$ which in coordinate basis can be written as:
\be
A_{\mu}=-\frac{1}{2m}\varepsilon_{\mu\nu\rho\sigma}\nabla^{\nu}B^{\rho\sigma}.
\ee
This means that the relationship between the polarization vector of $A$, $\epsilon^{(A)}$, and the polarization tensor of $B$, $\epsilon^{(B)}$, will be of the form:
\be
\epsilon^{(A)}_{\mu}\propto\frac{i}{m}\varepsilon_{\mu\nu\rho\sigma}p^{\nu}\epsilon^{(B)\rho\sigma},
\ee
where the overall constant can be found by requiring $\epsilon^{(A)}_{\mu}$ to be normalised ({\it i.e.} consistent with \eqref{eq:pol req}). This relation can be inverted by multiplying both sides by the $\varepsilon$ tensor and $p$, which using $p^2=-m^2$ and imposing normalisation condition gives \eqref{eq:pol map}.

\section{Double Copy of the 4-Point Scattering Amplitude in the Decoupling Limit}\label{sec:dl_mg}

We take the $\Lambda_3$ decoupling limit,
\be
m\rightarrow 0, \quad M_{pl}\rightarrow \infty, \quad \text{keeping }\Lambda_3=(m^2M_{pl})^{1/3} \text{ fixed},
\ee
of the full scattering amplitude obtained from double copy with external states arbitrary superpositions of $h$ and $\phi$ fields defined as: (setting the vectors to zero for simplicitly)
\begin{align}
\begin{split}
&\epsilon_{1\mu\nu}=\alpha_{T1}\epsilon^{2,+2}_{\mu\nu}(p_1)+\alpha_{T2}\epsilon^{2,-2}_{\mu\nu}(p_1)+\alpha_{T3}\epsilon^{2,+1}_{\mu\nu}(p_1)+\alpha_{T4}\epsilon^{2,-1}_{\mu\nu}(p_1)+\alpha_{T5}\epsilon^{2,0}_{\mu\nu}(p_1)+\alpha_S\epsilon^{0,0}_{\mu\nu}(p_1),\\
&\epsilon_{2\mu\nu}=\beta_{T1}\epsilon^{2,+2}_{\mu\nu}(p_2)+\beta_{T2}\epsilon^{2,-2}_{\mu\nu}(p_2)+\beta_{T3}\epsilon^{2,+1}_{\mu\nu}(p_2)+\beta_{T4}\epsilon^{2,-1}_{\mu\nu}(p_2)+\beta_{T5}\epsilon^{2,0}_{\mu\nu}(p_2)+\beta_S\epsilon^{0,0}_{\mu\nu}(p_2),\\
&\epsilon_{3\mu\nu}=\gamma_{T1}\epsilon^{2,+2}_{\mu\nu}(p_3)+\gamma_{T2}\epsilon^{2,-2}_{\mu\nu}(p_3)+\gamma_{T3}\epsilon^{2,+1}_{\mu\nu}(p_3)+\gamma_{T4}\epsilon^{2,-1}_{\mu\nu}(p_3)+\gamma_{T5}\epsilon^{2,0}_{\mu\nu}(p_3)+\gamma_S\epsilon^{0,0}_{\mu\nu}(p_3),\\
&\epsilon_{4\mu\nu}=\sigma_{T1}\epsilon^{2,+2}_{\mu\nu}(p_4)+\sigma_{T2}\epsilon^{2,-2}_{\mu\nu}(p_4)+\sigma_{T3}\epsilon^{2,+1}_{\mu\nu}(p_4)+\sigma_{T4}\epsilon^{2,-1}_{\mu\nu}(p_4)+\sigma_{T5}\epsilon^{2,0}_{\mu\nu}(p_4)+\sigma_S\epsilon^{0,0}_{\mu\nu}(p_4).
\end{split}
\end{align}
\allowdisplaybreaks
This gives the following amplitude:
\begin{align}
\begin{autobreak} \label{eq:fullamp}
 M_4\rightarrow
  i\frac{ s t u}{2304} 
 \Big(6 \alpha_{T3} \beta_{T3} \gamma_{S} \sigma_{S}
-6 \alpha_{T4} \beta_{T3} \gamma_{S} \sigma_{S}
-6 \alpha_{T3} \beta_{T4} \gamma_{S} \sigma_{S}
+6 \alpha_{T4} \beta_{T4} \gamma_{S} \sigma_{S}
+10 \alpha_{T5} \beta_{T5} \gamma_{S} \sigma_{S}
-6 \alpha_{T3} \beta_{S} \gamma_{T3} \sigma_{S}
+6 \alpha_{T4} \beta_{S} \gamma_{T3} \sigma_{S}
-6 \sqrt{2} \alpha_{T5} \beta_{T3} \gamma_{T3} \sigma_{S}
-6 \sqrt{2} \alpha_{T3} \beta_{T5} \gamma_{T3} \sigma_{S}
+6 \alpha_{T3} \beta_{S} \gamma_{T4} \sigma_{S}
-6 \alpha_{T4} \beta_{S} \gamma_{T4} \sigma_{S}
-6 \sqrt{2} \alpha_{T5} \beta_{T4} \gamma_{T4} \sigma_{S}
-6 \sqrt{2} \alpha_{T4} \beta_{T5} \gamma_{T4} \sigma_{S}
+10 \alpha_{T5} \beta_{S} \gamma_{T5} \sigma_{S}
-6 \sqrt{2} \alpha_{T4} \beta_{T3} \gamma_{T5} \sigma_{S}
-6 \sqrt{2} \alpha_{T3} \beta_{T4} \gamma_{T5} \sigma_{S}
-2 \sqrt{2} \alpha_{T5} \beta_{T5} \gamma_{T5} \sigma_{S}
+11 \alpha_{T5} \beta_{S} \gamma_{S} \sqrt{2} \sigma_{S}
+6 \alpha_{T5} \beta_{T4} \gamma_{T3} \sqrt{2} \sigma_{S}
+6 \alpha_{T4} \beta_{T5} \gamma_{T3} \sqrt{2} \sigma_{S}
+6 \alpha_{T5} \beta_{T3} \gamma_{T4} \sqrt{2} \sigma_{S}
+6 \alpha_{T3} \beta_{T5} \gamma_{T4} \sqrt{2} \sigma_{S}
+6 \alpha_{T3} \beta_{T3} \gamma_{T5} \sqrt{2} \sigma_{S}
+6 \alpha_{T4} \beta_{T4} \gamma_{T5} \sqrt{2} \sigma_{S}
+2 \alpha_{T2} \beta_{S} \gamma_{S} \sqrt{3} \sigma_{S}
+4 \alpha_{T5} \beta_{T5} \gamma_{T1} \sqrt{3} \sigma_{S}
+4 \alpha_{T5} \beta_{T5} \gamma_{T2} \sqrt{3} \sigma_{S}
+4 \alpha_{T5} \beta_{T1} \gamma_{T5} \sqrt{3} \sigma_{S}
+4 \alpha_{T5} \beta_{T2} \gamma_{T5} \sqrt{3} \sigma_{S}
+4 \alpha_{T2} \beta_{T5} \gamma_{T5} \sqrt{3} \sigma_{S}
+2 \alpha_{T5} \beta_{T1} \gamma_{S} \sqrt{6} \sigma_{S}
+2 \alpha_{T5} \beta_{T2} \gamma_{S} \sqrt{6} \sigma_{S}
+2 \alpha_{T2} \beta_{T5} \gamma_{S} \sqrt{6} \sigma_{S}
+2 \alpha_{T5} \beta_{S} \gamma_{T1} \sqrt{6} \sigma_{S}
+2 \alpha_{T5} \beta_{S} \gamma_{T2} \sqrt{6} \sigma_{S}
+2 \alpha_{T2} \beta_{S} \gamma_{T5} \sqrt{6} \sigma_{S}
-6 \alpha_{T3} \beta_{S} \gamma_{S} \sigma_{T3}
+6 \alpha_{T4} \beta_{S} \gamma_{S} \sigma_{T3}
-6 \sqrt{2} \alpha_{T5} \beta_{T3} \gamma_{S} \sigma_{T3}
-6 \sqrt{2} \alpha_{T3} \beta_{T5} \gamma_{S} \sigma_{T3}
+12 \alpha_{T5} \beta_{T5} \gamma_{T3} \sigma_{T3}
-6 \sqrt{2} \alpha_{T5} \beta_{S} \gamma_{T4} \sigma_{T3}
-12 \alpha_{T5} \beta_{T5} \gamma_{T4} \sigma_{T3}
-6 \sqrt{2} \alpha_{T3} \beta_{S} \gamma_{T5} \sigma_{T3}
-12 \alpha_{T5} \beta_{T3} \gamma_{T5} \sigma_{T3}
+12 \alpha_{T5} \beta_{T4} \gamma_{T5} \sigma_{T3}
-12 \alpha_{T3} \beta_{T5} \gamma_{T5} \sigma_{T3}
+12 \alpha_{T4} \beta_{T5} \gamma_{T5} \sigma_{T3}
+6 \alpha_{T3} \beta_{S} \gamma_{S} \sigma_{T4}
-6 \alpha_{T4} \beta_{S} \gamma_{S} \sigma_{T4}
-6 \sqrt{2} \alpha_{T5} \beta_{T4} \gamma_{S} \sigma_{T4}
-6 \sqrt{2} \alpha_{T4} \beta_{T5} \gamma_{S} \sigma_{T4}
-6 \sqrt{2} \alpha_{T5} \beta_{S} \gamma_{T3} \sigma_{T4}
-12 \alpha_{T5} \beta_{T5} \gamma_{T3} \sigma_{T4}
+12 \alpha_{T5} \beta_{T5} \gamma_{T4} \sigma_{T4}
-6 \sqrt{2} \alpha_{T4} \beta_{S} \gamma_{T5} \sigma_{T4}
+12 \alpha_{T5} \beta_{T3} \gamma_{T5} \sigma_{T4}
-12 \alpha_{T5} \beta_{T4} \gamma_{T5} \sigma_{T4}
+12 \alpha_{T3} \beta_{T5} \gamma_{T5} \sigma_{T4}
-12 \alpha_{T4} \beta_{T5} \gamma_{T5} \sigma_{T4}
+10 \alpha_{T5} \beta_{S} \gamma_{S} \sigma_{T5}
-6 \sqrt{2} \alpha_{T4} \beta_{T3} \gamma_{S} \sigma_{T5}
-6 \sqrt{2} \alpha_{T3} \beta_{T4} \gamma_{S} \sigma_{T5}
-2 \sqrt{2} \alpha_{T5} \beta_{T5} \gamma_{S} \sigma_{T5}
-6 \sqrt{2} \alpha_{T3} \beta_{S} \gamma_{T3} \sigma_{T5}
-12 \alpha_{T5} \beta_{T3} \gamma_{T3} \sigma_{T5}
+12 \alpha_{T5} \beta_{T4} \gamma_{T3} \sigma_{T5}
-12 \alpha_{T3} \beta_{T5} \gamma_{T3} \sigma_{T5}
+12 \alpha_{T4} \beta_{T5} \gamma_{T3} \sigma_{T5}
-6 \sqrt{2} \alpha_{T4} \beta_{S} \gamma_{T4} \sigma_{T5}
+12 \alpha_{T5} \beta_{T3} \gamma_{T4} \sigma_{T5}
-12 \alpha_{T5} \beta_{T4} \gamma_{T4} \sigma_{T5}
+12 \alpha_{T3} \beta_{T5} \gamma_{T4} \sigma_{T5}
-12 \alpha_{T4} \beta_{T5} \gamma_{T4} \sigma_{T5}
-2 \sqrt{2} \alpha_{T5} \beta_{S} \gamma_{T5} \sigma_{T5}
+12 \alpha_{T3} \beta_{T3} \gamma_{T5} \sigma_{T5}
-12 \alpha_{T4} \beta_{T3} \gamma_{T5} \sigma_{T5}
-12 \alpha_{T3} \beta_{T4} \gamma_{T5} \sigma_{T5}
+12 \alpha_{T4} \beta_{T4} \gamma_{T5} \sigma_{T5}
-28 \alpha_{T5} \beta_{T5} \gamma_{T5} \sigma_{T5}
-\alpha_{S} \Big(-11 \sqrt{2} \beta_{T5} \gamma_{S} \sigma_{S}
-2 \sqrt{6} \beta_{T5} \gamma_{T1} \sigma_{S}
-2 \sqrt{6} \beta_{T5} \gamma_{T2} \sigma_{S}
+6 \beta_{T3} \gamma_{T3} \sigma_{S}
-6 \beta_{T4} \gamma_{T3} \sigma_{S}
-6 \beta_{T3} \gamma_{T4} \sigma_{S}
+6 \beta_{T4} \gamma_{T4} \sigma_{S}
-2 \sqrt{6} \beta_{T2} \gamma_{T5} \sigma_{S}
-10 \beta_{T5} \gamma_{T5} \sigma_{S}
-2 \beta_{T2} \gamma_{S} \sqrt{3} \sigma_{S}
-2 \sqrt{6} \beta_{T5} \gamma_{S} \sigma_{T1}
-4 \sqrt{3} \beta_{T5} \gamma_{T5} \sigma_{T1}
-2 \sqrt{6} \beta_{T5} \gamma_{S} \sigma_{T2}
-4 \sqrt{3} \beta_{T5} \gamma_{T5} \sigma_{T2}
+6 \beta_{T3} \gamma_{S} \sigma_{T3}
-6 \beta_{T4} \gamma_{S} \sigma_{T3}
-6 \sqrt{2} \beta_{T5} \gamma_{T3} \sigma_{T3}
-6 \sqrt{2} \beta_{T4} \gamma_{T5} \sigma_{T3}
-6 \beta_{T3} \gamma_{S} \sigma_{T4}
+6 \beta_{T4} \gamma_{S} \sigma_{T4}
-6 \sqrt{2} \beta_{T5} \gamma_{T4} \sigma_{T4}
-6 \sqrt{2} \beta_{T3} \gamma_{T5} \sigma_{T4}
-2 \sqrt{6} \beta_{T2} \gamma_{S} \sigma_{T5}
-10 \beta_{T5} \gamma_{S} \sigma_{T5}
-4 \sqrt{3} \beta_{T5} \gamma_{T1} \sigma_{T5}
-4 \sqrt{3} \beta_{T5} \gamma_{T2} \sigma_{T5}
-6 \sqrt{2} \beta_{T4} \gamma_{T3} \sigma_{T5}
-6 \sqrt{2} \beta_{T3} \gamma_{T4} \sigma_{T5}
-4 \sqrt{3} \beta_{T2} \gamma_{T5} \sigma_{T5}
-\beta_{S} \Big(6 \gamma_{T3} \sigma_{T3}
-6 \gamma_{T4} \sigma_{T3}
-6 \gamma_{T3} \sigma_{T4}
+6 \gamma_{T4} \sigma_{T4}
+10 \gamma_{T5} \sigma_{T5}
-\gamma_{S} \Big(
-17 \sigma_{S}
-2 \sqrt{3} \sigma_{T1}
-2 \sqrt{3} \sigma_{T2}
-11 \sqrt{2} \sigma_{T5}\Big)
+11 \gamma_{T5} \sigma_{S} \sqrt{2}
+2 \gamma_{T1} \sigma_{S} \sqrt{3}
+2 \gamma_{T2} \sigma_{S} \sqrt{3}
+2 \gamma_{T5} \sigma_{T1} \sqrt{6}
+2 \gamma_{T5} \sigma_{T2} \sqrt{6}
+2 \gamma_{T1} \sigma_{T5} \sqrt{6}
+2 \gamma_{T2} \sigma_{T5} \sqrt{6}\Big)
+6 \beta_{T5} \gamma_{T4} \sigma_{T3} \sqrt{2}
+6 \beta_{T3} \gamma_{T5} \sigma_{T3} \sqrt{2}
+6 \beta_{T5} \gamma_{T3} \sigma_{T4} \sqrt{2}
+6 \beta_{T4} \gamma_{T5} \sigma_{T4} \sqrt{2}
+6 \beta_{T3} \gamma_{T3} \sigma_{T5} \sqrt{2}
+6 \beta_{T4} \gamma_{T4} \sigma_{T5} \sqrt{2}
+2 \beta_{T5} \gamma_{T5} \sigma_{T5} \sqrt{2}
+2 \beta_{T1} \Big(
-\gamma_{S} \sigma_{S}
-\sqrt{2} \gamma_{T5} \sigma_{S}
-\sqrt{2} \gamma_{S} \sigma_{T5}
-2 \gamma_{T5} \sigma_{T5}\Big) \sqrt{3}\Big)
+6 \alpha_{T5} \beta_{T4} \gamma_{S} \sigma_{T3} \sqrt{2}
+6 \alpha_{T4} \beta_{T5} \gamma_{S} \sigma_{T3} \sqrt{2}
+6 \alpha_{T5} \beta_{S} \gamma_{T3} \sigma_{T3} \sqrt{2}
+6 \alpha_{T4} \beta_{S} \gamma_{T5} \sigma_{T3} \sqrt{2}
+6 \alpha_{T5} \beta_{T3} \gamma_{S} \sigma_{T4} \sqrt{2}
+6 \alpha_{T3} \beta_{T5} \gamma_{S} \sigma_{T4} \sqrt{2}
+6 \alpha_{T5} \beta_{S} \gamma_{T4} \sigma_{T4} \sqrt{2}
+6 \alpha_{T3} \beta_{S} \gamma_{T5} \sigma_{T4} \sqrt{2}
+6 \alpha_{T3} \beta_{T3} \gamma_{S} \sigma_{T5} \sqrt{2}
+6 \alpha_{T4} \beta_{T4} \gamma_{S} \sigma_{T5} \sqrt{2}
+6 \alpha_{T4} \beta_{S} \gamma_{T3} \sigma_{T5} \sqrt{2}
+6 \alpha_{T3} \beta_{S} \gamma_{T4} \sigma_{T5} \sqrt{2}
+4 \alpha_{T5} \beta_{T5} \gamma_{S} \sigma_{T1} \sqrt{3}
+4 \alpha_{T5} \beta_{S} \gamma_{T5} \sigma_{T1} \sqrt{3}
+4 \alpha_{T5} \beta_{T5} \gamma_{S} \sigma_{T2} \sqrt{3}
+4 \alpha_{T5} \beta_{S} \gamma_{T5} \sigma_{T2} \sqrt{3}
+4 \alpha_{T5} \beta_{T1} \gamma_{S} \sigma_{T5} \sqrt{3}
+4 \alpha_{T5} \beta_{T2} \gamma_{S} \sigma_{T5} \sqrt{3}
+4 \alpha_{T2} \beta_{T5} \gamma_{S} \sigma_{T5} \sqrt{3}
+4 \alpha_{T5} \beta_{S} \gamma_{T1} \sigma_{T5} \sqrt{3}
+4 \alpha_{T5} \beta_{S} \gamma_{T2} \sigma_{T5} \sqrt{3}
+4 \alpha_{T2} \beta_{S} \gamma_{T5} \sigma_{T5} \sqrt{3}
+2 \alpha_{T1} \Big(\beta_{T5} \Big(2 \gamma_{T5} \sigma_{S}
+\gamma_{S} \sqrt{2} \sigma_{S}
+2 \gamma_{S} \sigma_{T5}
+2 \gamma_{T5} \sigma_{T5} \sqrt{2}\Big)
-\beta_{S} \Big(
-\gamma_{S} \sigma_{S}
-\sqrt{2} \gamma_{T5} \sigma_{S}
-\sqrt{2} \gamma_{S} \sigma_{T5}
-2 \gamma_{T5} \sigma_{T5}\Big)\Big) \sqrt{3}
+2 \alpha_{T5} \beta_{S} \gamma_{S} \sigma_{T1} \sqrt{6}
+4 \alpha_{T5} \beta_{T5} \gamma_{T5} \sigma_{T1} \sqrt{6}
+2 \alpha_{T5} \beta_{S} \gamma_{S} \sigma_{T2} \sqrt{6}
+4 \alpha_{T5} \beta_{T5} \gamma_{T5} \sigma_{T2} \sqrt{6}
+2 \alpha_{T2} \beta_{S} \gamma_{S} \sigma_{T5} \sqrt{6}
+4 \alpha_{T5} \beta_{T5} \gamma_{T1} \sigma_{T5} \sqrt{6}
+4 \alpha_{T5} \beta_{T5} \gamma_{T2} \sigma_{T5} \sqrt{6}
+4 \alpha_{T5} \beta_{T1} \gamma_{T5} \sigma_{T5} \sqrt{6}
+4 \alpha_{T5} \beta_{T2} \gamma_{T5} \sigma_{T5} \sqrt{6}
+4 \alpha_{T2} \beta_{T5} \gamma_{T5} \sigma_{T5} \sqrt{6}\Big).
\end{autobreak}
\end{align}

This amplitude simplifies considerable if we focus on scattering processes of the form $+2XXX$. We may easily see that $X$ can only be a scalar mode and this amplitude then takes the form
\be\label{eq:tensoramp}
M_4(+2 XXX) =  \frac{i stu}{96 \sqrt{6}} ( \beta_{T5}+\frac{1}{\sqrt{2}} \beta_S)( \gamma_{T5}+\frac{1}{\sqrt{2}} \gamma_S)( \sigma_{T5}+\frac{1}{\sqrt{2}} \sigma_S) \, .
\ee
The combination $ \beta_{T5}+\frac{1}{\sqrt{2}} \beta_S$ is precisely the combination of polarizations that picks out the helicity-0 squared term $\epsilon_0^{\mu} \epsilon_{0}^{\nu}$ 
\be\label{eq:comb}
\beta_{T5} \epsilon^{\mu\nu}_{2,0} + \beta_S   \epsilon^{\mu\nu}_{0,0} = \frac{2}{\sqrt{6}} ( \beta_{T5}+\frac{1}{\sqrt{2}} \beta_S)\epsilon_0^{\mu} \epsilon_{0}^{\nu} +\frac{1}{\sqrt{6}} (\beta_{T5}- \sqrt{2} \beta_{S}) (\epsilon_+^{\mu} \epsilon_-^{\nu}+ \epsilon_-^{\mu} \epsilon_+^{\nu})  \, .
\ee
Since the helicity $+2$ mode has polarization tensor $\epsilon_+^{\mu} \epsilon_{+}^{\nu}$ we recognize that $M_4(+2 XXX) $ is the double copy of the $+1000$ massive Yang-Mills amplitude and comes specifically from the $\Sigma \Sigma'$ contact term \eqref{eq:gravityDL}.

\section{Decoupling limit of massive Yang-Mills amplitude}\label{sec:dl ym}
In this section we derive the decoupling limit of the massive Yang-Mills amplitude which is expected to be the amplitude of NLSM, derive the kinematic factors and double copy it to show that we recover the 4 point amplitude of a special Galileon.  We also show that taking the decoupling limit and performing the double copy do not commute.
From \eqref{eq:An}, the 4-point amplitudes of massive Yang-Mills is expressed as:
\begin{equation}\label{eq: 4ptmym}
    A^{\text{mYM}}_{4}=\frac{m^2}{\Lambda^2}\left(\frac{c_s n_s}{s-m^2}+\frac{c_t n_t}{t-m^2}+\frac{c_u n_u}{u-m^2}\right),
\end{equation}
with the $n$'s given by \eqref{eq:ns}, \eqref{eq:nt} and \eqref{eq:nu}. By plugging the polarization vectors which are arbitrary superpositions of all helicities given as:
\begin{align}
\begin{split}
&\epsilon_{1\mu}=\alpha_{1}\epsilon^{+1}_{\mu}(p_1)+\alpha_{2}\epsilon^{-1}_{\mu}(p_1)+\alpha_{3}\epsilon^{0}_{\mu}(p_1),\\
&\epsilon_{2\mu}=\beta_{1}\epsilon^{+1}_{\mu}(p_2)+\beta_{2}\epsilon^{-1}_{\mu}(p_2)+\beta_{3}\epsilon^{0}_{\mu}(p_2),\\
&\epsilon_{3\mu}=\gamma_{1}\epsilon^{+1}_{\mu}(p_3)+\gamma_{2}\epsilon^{-1}_{\mu}(p_3)+\gamma_{3}\epsilon^{0}_{\mu}(p_3),\\
&\epsilon_{4\mu}=\sigma_{1}\epsilon^{+1}_{\mu}(p_4)+\sigma_{2}\epsilon^{-1}_{\mu}(p_4)+\sigma_{3}\epsilon^{0}_{\mu}(p_4),
\end{split}
\end{align}
 and four momenta into the $n$'s, they can be rearranged in the following form (as mentioned in \eqref{kinematic}):

\be\label{kinematic2}
n_s = \frac{s-m^2}{m^3} \Sigma(s,t,u) + \frac{1}{m^2} \hat n_s, \quad n_t = \frac{t-m^2}{m^3} \Sigma(s,t,u) + \frac{1}{m^2} \hat n_t, \quad n_u = \frac{u-m^2}{m^3} \Sigma(s,t,u) + \frac{1}{m^2} \hat n_u \, ,
\ee
with $n_s+n_t+n_u=0$ and $\hat n_s+\hat n_t+\hat n_u=-m \Sigma$. The explicit expressions for the $\hat n$'s and $\Sigma(s,t,u)$ are given in \eqref{eq: Sigma} and \eqref{eq: nshat} \eqref{eq: nthat} \eqref{eq: nuhat}. The amplitude can be written as,

\begin{align}
 A^{\text{mYM}}_{4}=&\frac{m^2}{\Lambda^2}\left(\frac{c_s n_s}{s-m^2}+\frac{c_t n_t}{t-m^2}+\frac{c_u n_u}{u-m^2}\right),\\
 =&  \frac{1}{\Lambda^2}\left(\frac{c_s \hat n_s}{s-m^2}+\frac{c_t \hat n_t}{t-m^2}+\frac{c_u \hat n_u}{u-m^2}\right)+\frac{1}{m\Lambda^2}\Sigma(s,t,u)\left(c_s+c_t+c_u\right),
\end{align}
and as mentioned in the introduction, the last term which seems at first ill defined in the decoupling limit $m\rightarrow 0$, $\Lambda$ fixed, is zero by virtue of Jacobi identity. Focusing on the non-zero term, the amplitude in the dcoupling limit is as follows:
\begin{align}\label{nlsm am}
 \begin{split}
    A^{\text{DL}}_{4}&=\lim_{m\rightarrow0,\;\Lambda \text{fixed}}\frac{1}{\Lambda^2}A^{\text{mYM}},\\
     &=-i\frac{1}{12\Lambda^2}\Bigg(c_{s} (t-u) + c_{t}(u-s)+c_{u}(s-t)\Bigg)\alpha_3\beta_3\sigma_3\gamma_3.
\end{split}
\end{align}

We see that only helicity-0 polarization states remain interacting in this decoupling limit. The kinematic factors of this amplitude are,
\begin{equation}
    n_s=-\frac{i s}{12}(t-u),\;\quad n_t=-\frac{i t}{12}(u-s),\;\quad n_u=-\frac{i u}{12}(s-t).
\end{equation}
Note that in this limit we have $s+t+u=0$ and can see that the colour-kinematics duality is satisfied.\\
Using the kinematic factors of this amplitude we double copy it and obtain the following:
\begin{equation}
    A^{DC}=i \frac{\alpha _3^2 \beta _3^2 \gamma _3^2 \sigma _3^2 }{16 \Lambda_{3}^6}stu,
\end{equation}
which is equal to the scattering amplitude of a galileon theory.\\

It seems that we could have defined the kinematic factors of the full massive Yang-Mills theory without the $1/m^3$ terms in \eqref{kinematic2} since they cancel in the full amplitude. However without them the colour-kinematics duality is not satisfied. This is in contrast to the massless double copy where at four-points any representation of kinematic factors satisfy this duality. If we tried to double copy, {\it i.e.}
\be\label{eq:gravityDL2}
\frac{1}{\mpl^2} \(\frac{n_sn_s'}{s-m^2}+\frac{n_tn_t'}{t-m^2}+\frac{n_un_u'}{u-m^2} \)= \frac{-\Sigma \Sigma'}{\Lambda_3^6} +\frac{1}{\Lambda_3^6} \(\frac{\hat n_s \hat n_s'}{s-m^2}+\frac{\hat n_t \hat n_t'}{t-m^2}+\frac{\hat n_u \hat n_u'}{u-m^2} \) \,,
\ee
without using  $\Sigma(s,t,u)$, we would have obtained a theory whose $\Lambda_3$ decoupling limit is the special galileon because only $\hat n$ terms could have contributed to the double copy amplitude, {\it i.e.} we would have obtained
\begin{equation}\label{sGl}
    A_{\hat n^2}^{DC}=\frac{i}{M_{pl}^2}\sum_{i=1}^{3}\frac{\hat n_{i}\hat n'_{i}}{m^4s_i}=i\frac{\alpha _3^2 \beta _3^2 \gamma _3^2 \sigma _3^2 }{16 \Lambda_{3}^6}stu,
\end{equation}
where $i=1,2,3$ labels $s, t ,u$ respectively. However, in our case, when we square $\Sigma(s,t,u)$ , they sum to a $1/m^4$ contribution to the double copy scattering amplitude 
\begin{align}\label{no sGl}
\begin{split}
     A_{\Sigma^2(s,t,u)}^{DC}= \\&i\frac{\left(\alpha _3 \beta _3 \gamma _3 \left(\sigma _2-\sigma _1\right)+\alpha _3 \beta _3 \left(\gamma _2-\gamma _1\right) \sigma _3+\gamma _3 \sigma _3 \left(\alpha _3 \left(\beta _1-\beta _2\right)+\left(\alpha _1-\alpha _2\right) \beta _3\right)\right){}^2}{8 \Lambda_{3}^6}stu,
\end{split}
\end{align}
 which contains helicity $\pm1$ polarizations and the decoupling limit of the resulting theory is not the double copy of the decoupling limit of the massive Yang-Mills, {\it i.e.} the operations of taking decoupling limit and performing double copy do not commute. \\
 
The explicit expressions for $\Sigma(s,t,u)$, $\hat n_s$, $\hat n_t$, $\hat n_u$ are given below:

 \be\label{eq: Sigma}
 \Sigma(s,t,u)=i\frac{\sqrt{s t u} \left(\alpha _3 \beta _3 \gamma _3 \left(\sigma _1-\sigma _2\right)+\alpha _3 \beta _3 \left(\gamma _1-\gamma _2\right) \sigma _3+\gamma _3 \sigma _3 \left(\alpha _3 \left(\beta _2-\beta _1\right)+\left(\alpha _2-\alpha _1\right) \beta _3\right)\right)}{2 \sqrt{2}}
\ee

\begin{align}\label{eq: nshat}
 \begin{autobreak}
\hat n_{s}=
-\frac{i}{4 \Big(4 m^2-s\big)} 
\big(16 \big(\alpha _2 \big(
-\beta _2 \gamma _1 \sigma _1
+\beta _2 \gamma _2 \sigma _2
-\beta _3 \gamma _3 \sigma _2
+\beta _3 \gamma _1 \sigma _3\big)
+\alpha _1 \big(\beta _1 \gamma _1 \sigma _1
-\beta _3 \gamma _3 \sigma _1
-\beta _1 \gamma _2 \sigma _2
+\beta _3 \gamma _2 \sigma _3\big)
+\alpha _3 \big(\beta _2 \gamma _3 \sigma _1
+\beta _1 \gamma _3 \sigma _2
-\beta _1 \gamma _1 \sigma _3
-\beta _2 \gamma _2 \sigma _3\big)\big) m^6
-4 \big(4 u \big(\alpha _1 \beta _1
+\alpha _2 \beta _2
-\alpha _3 \beta _3\big) \big(\gamma _1 \sigma _1
+\gamma _2 \sigma _2
-\gamma _3 \sigma _3\big)
+t \big(
-\alpha _3 \big(\beta _1
-\beta _2\big) \big(\gamma _3 \big(\sigma _1
-\sigma _2\big)
+\big(\gamma _1
-\gamma _2\big) \sigma _3\big)
+\alpha _2 \beta _3 \big(\gamma _3 \big(\sigma _1
-\sigma _2\big)
+\big(\gamma _1
-\gamma _2\big) \sigma _3\big)
+4 \alpha _3 \beta _3 \big(\gamma _1 \sigma _1
+\gamma _2 \sigma _2
-\gamma _3 \sigma _3\big)
-2 \alpha _2 \beta _2 \big(3 \gamma _1 \sigma _1
+\gamma _2 \sigma _2
-2 \gamma _3 \sigma _3\big)
+\alpha _1 \big(\beta _3 \gamma _3 \big(\sigma _2
-\sigma _1\big)
+\beta _3 \big(\gamma _2
-\gamma _1\big) \sigma _3
-2 \beta _1 \big(\gamma _1 \sigma _1
+3 \gamma _2 \sigma _2
-2 \gamma _3 \sigma _3\big)\big)\big)
+s \big(\alpha _1 \big(5 \beta _1 \gamma _1 \sigma _1
+9 \beta _3 \gamma _3 \sigma _1
-5 \beta _1 \gamma _2 \sigma _2
-11 \beta _3 \gamma _2 \sigma _3
-16 \beta _1 \gamma _3 \sigma _3\big)
+\alpha _2 \big(
-5 \beta _2 \gamma _1 \sigma _1
+5 \beta _2 \gamma _2 \sigma _2
+9 \beta _3 \gamma _3 \sigma _2
-11 \beta _3 \gamma _1 \sigma _3
-16 \beta _2 \gamma _3 \sigma _3\big)
+\alpha _3 \big(
-11 \beta _2 \gamma _3 \sigma _1
-11 \beta _1 \gamma _3 \sigma _2
+9 \beta _1 \gamma _1 \sigma _3
+9 \beta _2 \gamma _2 \sigma _3
-4 \beta _3 \big(4 \gamma _1 \sigma _1
+4 \gamma _2 \sigma _2
-9 \gamma _3 \sigma _3\big)\big)\big)\big) m^4
+2 \sqrt{2} \sqrt{s t u} \big(\alpha _2 \beta _2 \big(
-9 \gamma _3 \sigma _1
+7 \gamma _3 \sigma _2
-9 \gamma _1 \sigma _3
+7 \gamma _2 \sigma _3\big)
+\alpha _1 \beta _1 \big(
-7 \gamma _3 \sigma _1
+9 \gamma _3 \sigma _2
-7 \gamma _1 \sigma _3
+9 \gamma _2 \sigma _3\big)
+\alpha _2 \beta _3 \big(
-9 \gamma _1 \sigma _1
-7 \gamma _2 \sigma _2
+12 \gamma _3 \sigma _3\big)
+\alpha _1 \beta _3 \big(7 \gamma _1 \sigma _1
+9 \gamma _2 \sigma _2
-12 \gamma _3 \sigma _3\big)
+\alpha _3 \big(\beta _2 \big(
-9 \gamma _1 \sigma _1
-7 \gamma _2 \sigma _2
+12 \gamma _3 \sigma _3\big)
+12 \beta _3 \big(\gamma _3 \sigma _1
-\gamma _3 \sigma _2
+\gamma _1 \sigma _3
-\gamma _2 \sigma _3\big)
+\beta _1 \big(7 \gamma _1 \sigma _1
+9 \gamma _2 \sigma _2
-12 \gamma _3 \sigma _3\big)\big)\big) m^3
+2 s \big(t \big(\alpha _3 \beta _2 \big(
-6 \gamma _3 \sigma _1
+5 \gamma _3 \sigma _2
-6 \gamma _1 \sigma _3
+5 \gamma _2 \sigma _3\big)
+\alpha _2 \beta _3 \big(
-6 \gamma _3 \sigma _1
+5 \gamma _3 \sigma _2
-6 \gamma _1 \sigma _3
+5 \gamma _2 \sigma _3\big)
-2 \alpha _2 \beta _2 \big(3 \gamma _1 \sigma _1
-\gamma _2 \sigma _2
+5 \gamma _3 \sigma _3\big)
+\alpha _3 \beta _1 \big(5 \gamma _3 \sigma _1
-6 \gamma _3 \sigma _2
+5 \gamma _1 \sigma _3
-6 \gamma _2 \sigma _3\big)
+\alpha _1 \beta _3 \big(5 \gamma _3 \sigma _1
-6 \gamma _3 \sigma _2
+5 \gamma _1 \sigma _3
-6 \gamma _2 \sigma _3\big)
+2 \alpha _1 \beta _1 \big(\gamma _1 \sigma _1
-3 \gamma _2 \sigma _2
-5 \gamma _3 \sigma _3\big)
-2 \alpha _3 \beta _3 \big(5 \gamma _1 \sigma _1
+5 \gamma _2 \sigma _2
-13 \gamma _3 \sigma _3\big)\big)
+2 u \big(\alpha _3 \beta _3 \big(
-3 \gamma _1 \sigma _1
-3 \gamma _2 \sigma _2
+5 \gamma _3 \sigma _3\big)
+\alpha _1 \beta _1 \big(\gamma _1 \sigma _1
+\gamma _2 \sigma _2
-3 \gamma _3 \sigma _3\big)
+\alpha _2 \beta _2 \big(\gamma _1 \sigma _1
+\gamma _2 \sigma _2
-3 \gamma _3 \sigma _3\big)\big)
+s \big(\alpha _2 \big(\beta _3 \big(9 \gamma _3 \sigma _2
-14 \gamma _1 \sigma _3\big)
-2 \beta _2 \big(\gamma _1 \sigma _1
-\gamma _2 \sigma _2
+8 \gamma _3 \sigma _3\big)\big)
+\alpha _1 \big(\beta _3 \big(9 \gamma _3 \sigma _1
-14 \gamma _2 \sigma _3\big)
+2 \beta _1 \big(\gamma _1 \sigma _1
-\gamma _2 \sigma _2
-8 \gamma _3 \sigma _3\big)\big)
+\alpha _3 \big(
-14 \beta _2 \gamma _3 \sigma _1
-14 \beta _1 \gamma _3 \sigma _2
+9 \beta _1 \gamma _1 \sigma _3
+9 \beta _2 \gamma _2 \sigma _3
-4 \beta _3 \big(4 \gamma _1 \sigma _1
+4 \gamma _2 \sigma _2
-11 \gamma _3 \sigma _3\big)\big)\big)\big) m^2
-2 \sqrt{2} s \sqrt{s t u} \big(\alpha _2 \beta _2 \big(\gamma _3 \big(\sigma _2
-3 \sigma _1\big)
+\big(\gamma _2
-3 \gamma _1\big) \sigma _3\big)
-\alpha _1 \beta _1 \big(\gamma _3 \big(\sigma _1
-3 \sigma _2\big)
+\big(\gamma _1
-3 \gamma _2\big) \sigma _3\big)
-\alpha _2 \beta _3 \big(3 \gamma _1 \sigma _1
+\gamma _2 \sigma _2
-3 \gamma _3 \sigma _3\big)
+\alpha _1 \beta _3 \big(\gamma _1 \sigma _1
+3 \gamma _2 \sigma _2
-3 \gamma _3 \sigma _3\big)
+\alpha _3 \big(3 \beta _3 \big(\gamma _3 \sigma _1
-\gamma _3 \sigma _2
+\gamma _1 \sigma _3
-\gamma _2 \sigma _3\big)
-\beta _2 \big(3 \gamma _1 \sigma _1
+\gamma _2 \sigma _2
-3 \gamma _3 \sigma _3\big)
+\beta _1 \big(\gamma _1 \sigma _1
+3 \gamma _2 \sigma _2
-3 \gamma _3 \sigma _3\big)\big)\big) m
+4 s^3 \alpha _3 \beta _3 \gamma _3 \sigma _3
+s^2 \big(s \big(
-2 \alpha _2 \beta _3 \gamma _3 \sigma _2
+4 \alpha _2 \beta _3 \gamma _1 \sigma _3
+4 \alpha _2 \beta _2 \gamma _3 \sigma _3
+\alpha _1 \big(
-2 \beta _3 \gamma _3 \sigma _1
+4 \beta _3 \gamma _2 \sigma _3
+4 \beta _1 \gamma _3 \sigma _3\big)
+\alpha _3 \big(4 \beta _2 \gamma _3 \sigma _1
+4 \beta _1 \gamma _3 \sigma _2
-2 \beta _1 \gamma _1 \sigma _3
-2 \beta _2 \gamma _2 \sigma _3
+\beta _3 \big(4 \gamma _1 \sigma _1
+4 \gamma _2 \sigma _2
-17 \gamma _3 \sigma _3\big)\big)\big)
+2 \big(u \big(\big(\alpha _1 \beta _1
+\alpha _2 \beta _2\big) \gamma _3 \sigma _3
+\alpha _3 \beta _3 \big(\gamma _1 \sigma _1
+\gamma _2 \sigma _2
-4 \gamma _3 \sigma _3\big)\big)
+t \big(3 \alpha _1 \beta _1 \gamma _3 \sigma _3
+3 \alpha _2 \beta _2 \gamma _3 \sigma _3
-\alpha _1 \beta _3 \big(\gamma _3 \big(\sigma _1
-2 \sigma _2\big)
+\big(\gamma _1
-2 \gamma _2\big) \sigma _3\big)
+\alpha _2 \beta _3 \big(2 \gamma _3 \sigma _1
-\gamma _3 \sigma _2
+2 \gamma _1 \sigma _3
-\gamma _2 \sigma _3\big)
-\alpha _3 \big(\beta _2 \big(
-2 \gamma _3 \sigma _1
+\gamma _3 \sigma _2
-2 \gamma _1 \sigma _3
+\gamma _2 \sigma _3\big)
+\beta _1 \big(\gamma _3 \sigma _1
-2 \gamma _3 \sigma _2
+\gamma _1 \sigma _3
-2 \gamma _2 \sigma _3\big)
-3 \beta _3 \big(\gamma _1 \sigma _1
+\gamma _2 \sigma _2
-3 \gamma _3 \sigma _3\big)\big)\big)\big)\big)\Big)
\end{autobreak}
\end{align}

\begin{align}\label{eq: nthat}
\begin{autobreak}
   \hat n_{t}=
\frac{i}{4\left(s-4 m^2\right)^2}
\Big(
-64 \big(
-\alpha _3 \big(\beta _3 \gamma _1 \sigma _1
+4 \beta _2 \gamma _3 \sigma _1
+\beta _3 \gamma _2 \sigma _2
+4 \beta _1 \gamma _3 \sigma _2\big)
+\alpha _3 \big(\beta _1 \gamma _1
+\beta _2 \gamma _2
+4 \beta _3 \gamma _3\big) \sigma _3
+\alpha _1 \big(3 \beta _2 \gamma _2 \sigma _1
+\beta _3 \gamma _3 \sigma _1
+5 \beta _1 \gamma _2 \sigma _2
-4 \beta _3 \gamma _2 \sigma _3
-\beta _1 \gamma _3 \sigma _3\big)
+\alpha _2 \big(5 \beta _2 \gamma _1 \sigma _1
+3 \beta _1 \gamma _1 \sigma _2
+\beta _3 \gamma _3 \sigma _2
-4 \beta _3 \gamma _1 \sigma _3
-\beta _2 \gamma _3 \sigma _3\big)\big) m^8
-16 \big(t \big(\alpha _3 \big(4 \gamma _3 \big(\beta _1 \sigma _1
+\beta _2 \sigma _2\big)
+\big(3 \beta _1 \gamma _1
-7 \beta _2 \gamma _1
-7 \beta _1 \gamma _2
+3 \beta _2 \gamma _2\big) \sigma _3
-4 \beta _3 \big(
-2 \gamma _2 \sigma _1
+\gamma _2 \sigma _2
+\gamma _1 \big(\sigma _1
-2 \sigma _2\big)
+\gamma _3 \sigma _3\big)\big)
-\alpha _2 \big(\beta _2 \big(
-16 \gamma _1 \sigma _1
+5 \gamma _2 \sigma _1
+5 \gamma _1 \sigma _2
-2 \gamma _2 \sigma _2
+4 \gamma _3 \sigma _3\big)
+\beta _3 \big(7 \gamma _3 \sigma _1
-3 \gamma _3 \sigma _2
-4 \gamma _2 \sigma _3\big)
+\beta _1 \big(5 \gamma _1 \sigma _1
+6 \gamma _1 \sigma _2
+5 \gamma _2 \sigma _2
-8 \gamma _3 \sigma _3\big)\big)
+\alpha _1 \big(\beta _3 \big(3 \gamma _3 \sigma _1
-7 \gamma _3 \sigma _2
+4 \gamma _1 \sigma _3\big)
+\beta _1 \big(2 \gamma _1 \sigma _1
-5 \gamma _2 \sigma _1
-5 \gamma _1 \sigma _2
+16 \gamma _2 \sigma _2
-4 \gamma _3 \sigma _3\big)
-\beta _2 \big(5 \gamma _1 \sigma _1
+6 \gamma _2 \sigma _1
+5 \gamma _2 \sigma _2
-8 \gamma _3 \sigma _3\big)\big)\big)
-2 s \big(
-2 \alpha _3 \big(\beta _3 \gamma _1 \sigma _1
+4 \beta _2 \gamma _3 \sigma _1
+\beta _3 \gamma _2 \sigma _2
+4 \beta _1 \gamma _3 \sigma _2\big)
+2 \alpha _3 \big(\beta _1 \gamma _1
+\beta _2 \gamma _2
+4 \beta _3 \gamma _3\big) \sigma _3
+\alpha _1 \big(7 \beta _2 \gamma _2 \sigma _1
+2 \beta _3 \gamma _3 \sigma _1
+9 \beta _1 \gamma _2 \sigma _2
-8 \beta _3 \gamma _2 \sigma _3
-2 \beta _1 \gamma _3 \sigma _3\big)
+\alpha _2 \big(9 \beta _2 \gamma _1 \sigma _1
+7 \beta _1 \gamma _1 \sigma _2
+2 \beta _3 \gamma _3 \sigma _2
-8 \beta _3 \gamma _1 \sigma _3
-2 \beta _2 \gamma _3 \sigma _3\big)\big)\big) m^6
-8 \sqrt{2} \sqrt{s t u} \big(\alpha _1 \big(\beta _3 \gamma _1 \sigma _1
-5 \beta _3 \gamma _2 \sigma _1
-\beta _1 \gamma _3 \sigma _1
+5 \beta _2 \gamma _3 \sigma _1
+4 \beta _3 \gamma _2 \sigma _2
+4 \beta _1 \gamma _3 \sigma _2
-\big(\beta _1 \gamma _1
-4 \beta _1 \gamma _2
+5 \beta _2 \gamma _2
+5 \beta _3 \gamma _3\big) \sigma _3\big)
+\alpha _3 \big(\beta _2 \big(
-4 \gamma _1 \sigma _1
+5 \gamma _2 \sigma _1
-\gamma _2 \sigma _2
+5 \gamma _3 \sigma _3\big)
+5 \beta _3 \big(\gamma _3 \sigma _1
-\gamma _3 \sigma _2
+\gamma _1 \sigma _3
-\gamma _2 \sigma _3\big)
+\beta _1 \big(4 \gamma _2 \sigma _2
+\gamma _1 \big(\sigma _1
-5 \sigma _2\big)
-5 \gamma _3 \sigma _3\big)\big)
+\alpha _2 \big(
-5 \beta _1 \gamma _3 \sigma _2
+5 \beta _1 \gamma _1 \sigma _3
+\beta _2 \big(
-4 \gamma _3 \sigma _1
+\gamma _3 \sigma _2
-4 \gamma _1 \sigma _3
+\gamma _2 \sigma _3\big)
+\beta _3 \big(
-4 \gamma _1 \sigma _1
+5 \gamma _1 \sigma _2
-\gamma _2 \sigma _2
+5 \gamma _3 \sigma _3\big)\big)\big) m^5
-4 \big(\big(
-5 \alpha _3 \big(\beta _3 \gamma _1 \sigma _1
+4 \beta _2 \gamma _3 \sigma _1
+\beta _3 \gamma _2 \sigma _2
+4 \beta _1 \gamma _3 \sigma _2\big)
+5 \alpha _3 \big(\beta _1 \gamma _1
+\beta _2 \gamma _2
+4 \beta _3 \gamma _3\big) \sigma _3
+\alpha _1 \big(19 \beta _2 \gamma _2 \sigma _1
+5 \beta _3 \gamma _3 \sigma _1
+21 \beta _1 \gamma _2 \sigma _2
-5 \big(4 \beta _3 \gamma _2
+\beta _1 \gamma _3\big) \sigma _3\big)
+\alpha _2 \big(21 \beta _2 \gamma _1 \sigma _1
+19 \beta _1 \gamma _1 \sigma _2
+5 \beta _3 \gamma _3 \sigma _2
-5 \big(4 \beta _3 \gamma _1
+\beta _2 \gamma _3\big) \sigma _3\big)\big) s^2
+t \big(
-2 \alpha _3 \big(\beta _2 \gamma _3 \big(15 \sigma _1
+\sigma _2\big)
+\beta _1 \gamma _3 \big(\sigma _1
+15 \sigma _2\big)
+\beta _3 \big(5 \gamma _1 \sigma _1
+3 \gamma _2 \sigma _1
+3 \gamma _1 \sigma _2
+5 \gamma _2 \sigma _2\big)\big)
+\alpha _3 \big(13 \beta _1 \gamma _1
+5 \beta _2 \gamma _1
+5 \beta _1 \gamma _2
+13 \beta _2 \gamma _2
+52 \beta _3 \gamma _3\big) \sigma _3
+\alpha _1 \big(
-2 \beta _1 \gamma _1 \sigma _1
+5 \beta _2 \gamma _1 \sigma _1
+5 \beta _1 \gamma _2 \sigma _1
+22 \beta _2 \gamma _2 \sigma _1
+13 \beta _3 \gamma _3 \sigma _1
+5 \beta _1 \gamma _1 \sigma _2
-24 \beta _1 \gamma _2 \sigma _2
+5 \beta _2 \gamma _2 \sigma _2
+5 \beta _3 \gamma _3 \sigma _2
-2 \big(\beta _3 \gamma _1
+15 \beta _3 \gamma _2
+5 \beta _1 \gamma _3
+3 \beta _2 \gamma _3\big) \sigma _3\big)
+\alpha _2 \big(5 \beta _1 \gamma _1 \sigma _1
-24 \beta _2 \gamma _1 \sigma _1
+5 \beta _2 \gamma _2 \sigma _1
+5 \beta _3 \gamma _3 \sigma _1
+22 \beta _1 \gamma _1 \sigma _2
+5 \beta _2 \gamma _1 \sigma _2
+5 \beta _1 \gamma _2 \sigma _2
-2 \beta _2 \gamma _2 \sigma _2
+13 \beta _3 \gamma _3 \sigma _2
-2 \big(15 \beta _3 \gamma _1
+\beta _3 \gamma _2
+3 \beta _1 \gamma _3
+5 \beta _2 \gamma _3\big) \sigma _3\big)\big) s
+t^2 \big(8 \alpha _3 \beta _3 \big(\gamma _1
-\gamma _2\big) \big(\sigma _1
-\sigma _2\big)
-4 \alpha _3 \big(\beta _1
-\beta _2\big) \big(\gamma _3 \big(\sigma _1
-\sigma _2\big)
+\big(\gamma _1
-\gamma _2\big) \sigma _3\big)
+\alpha _2 \big(\beta _2 \big(
-21 \gamma _1 \sigma _1
+5 \gamma _2 \sigma _1
+5 \gamma _1 \sigma _2
-5 \gamma _2 \sigma _2
+8 \gamma _3 \sigma _3\big)
+4 \beta _3 \big(\gamma _3 \sigma _1
-\gamma _3 \sigma _2
+\gamma _1 \sigma _3
-\gamma _2 \sigma _3\big)
+\beta _1 \big(5 \gamma _1 \sigma _1
+3 \gamma _2 \sigma _1
+3 \gamma _1 \sigma _2
+5 \gamma _2 \sigma _2
-8 \gamma _3 \sigma _3\big)\big)
+\alpha _1 \big(4 \beta _3 \big(
-\gamma _3 \sigma _1
+\gamma _3 \sigma _2
-\gamma _1 \sigma _3
+\gamma _2 \sigma _3\big)
+\beta _1 \big(5 \gamma _2 \sigma _1
-21 \gamma _2 \sigma _2
+5 \gamma _1 \big(\sigma _2
-\sigma _1\big)
+8 \gamma _3 \sigma _3\big)
+\beta _2 \big(5 \gamma _1 \sigma _1
+3 \gamma _2 \sigma _1
+3 \gamma _1 \sigma _2
+5 \gamma _2 \sigma _2
-8 \gamma _3 \sigma _3\big)\big)\big)\big) m^4
+2 \sqrt{2} \sqrt{s t u} \big(s \big(\alpha _1 \big(\beta _3 \gamma _1 \sigma _1
-5 \beta _3 \gamma _2 \sigma _1
-\beta _1 \gamma _3 \sigma _1
+5 \beta _2 \gamma _3 \sigma _1
+4 \beta _3 \gamma _2 \sigma _2
+4 \beta _1 \gamma _3 \sigma _2
-\big(\beta _1 \gamma _1
-4 \beta _1 \gamma _2
+5 \beta _2 \gamma _2
+5 \beta _3 \gamma _3\big) \sigma _3\big)
+\alpha _3 \big(\beta _2 \big(
-4 \gamma _1 \sigma _1
+5 \gamma _2 \sigma _1
-\gamma _2 \sigma _2
+5 \gamma _3 \sigma _3\big)
+5 \beta _3 \big(\gamma _3 \sigma _1
-\gamma _3 \sigma _2
+\gamma _1 \sigma _3
-\gamma _2 \sigma _3\big)
+\beta _1 \big(4 \gamma _2 \sigma _2
+\gamma _1 \big(\sigma _1
-5 \sigma _2\big)
-5 \gamma _3 \sigma _3\big)\big)
+\alpha _2 \big(
-5 \beta _1 \gamma _3 \sigma _2
+5 \beta _1 \gamma _1 \sigma _3
+\beta _2 \big(
-4 \gamma _3 \sigma _1
+\gamma _3 \sigma _2
-4 \gamma _1 \sigma _3
+\gamma _2 \sigma _3\big)
+\beta _3 \big(
-4 \gamma _1 \sigma _1
+5 \gamma _1 \sigma _2
-\gamma _2 \sigma _2
+5 \gamma _3 \sigma _3\big)\big)\big)
+t \big(\alpha _3 \big(\beta _2 \big(
-17 \gamma _1 \sigma _1
+5 \gamma _2 \sigma _1
+5 \gamma _1 \sigma _2
-9 \gamma _2 \sigma _2
+24 \gamma _3 \sigma _3\big)
+24 \beta _3 \big(\gamma _3 \sigma _1
-\gamma _3 \sigma _2
+\gamma _1 \sigma _3
-\gamma _2 \sigma _3\big)
+\beta _1 \big(9 \gamma _1 \sigma _1
-5 \gamma _2 \sigma _1
-5 \gamma _1 \sigma _2
+17 \gamma _2 \sigma _2
-24 \gamma _3 \sigma _3\big)\big)
+\alpha _2 \big(\beta _2 \big(
-17 \gamma _3 \sigma _1
+9 \gamma _3 \sigma _2
-17 \gamma _1 \sigma _3
+9 \gamma _2 \sigma _3\big)
+\beta _3 \big(
-17 \gamma _1 \sigma _1
+5 \gamma _2 \sigma _1
+5 \gamma _1 \sigma _2
-9 \gamma _2 \sigma _2
+24 \gamma _3 \sigma _3\big)
+5 \beta _1 \big(\gamma _3 \sigma _1
-\gamma _3 \sigma _2
+\gamma _1 \sigma _3
-\gamma _2 \sigma _3\big)\big)
+\alpha _1 \big(\beta _1 \big(
-9 \gamma _3 \sigma _1
+17 \gamma _3 \sigma _2
-9 \gamma _1 \sigma _3
+17 \gamma _2 \sigma _3\big)
+5 \beta _2 \big(\gamma _3 \sigma _1
-\gamma _3 \sigma _2
+\gamma _1 \sigma _3
-\gamma _2 \sigma _3\big)
+\beta _3 \big(9 \gamma _1 \sigma _1
-5 \gamma _2 \sigma _1
-5 \gamma _1 \sigma _2
+17 \gamma _2 \sigma _2
-24 \gamma _3 \sigma _3\big)\big)\big)\big) m^3
+2 s \big(\big(
-\alpha _3 \big(\beta _3 \gamma _1 \sigma _1
+4 \beta _2 \gamma _3 \sigma _1
+\beta _3 \gamma _2 \sigma _2
+4 \beta _1 \gamma _3 \sigma _2\big)
+\alpha _3 \big(\beta _1 \gamma _1
+\beta _2 \gamma _2
+4 \beta _3 \gamma _3\big) \sigma _3
+\alpha _1 \big(4 \beta _2 \gamma _2 \sigma _1
+\beta _3 \gamma _3 \sigma _1
+4 \beta _1 \gamma _2 \sigma _2
-4 \beta _3 \gamma _2 \sigma _3
-\beta _1 \gamma _3 \sigma _3\big)
+\alpha _2 \big(4 \beta _2 \gamma _1 \sigma _1
+4 \beta _1 \gamma _1 \sigma _2
+\beta _3 \gamma _3 \sigma _2
-4 \beta _3 \gamma _1 \sigma _3
-\beta _2 \gamma _3 \sigma _3\big)\big) s^2
+t \big(\alpha _2 \big(
-4 \beta _2 \gamma _1 \sigma _1
-\beta _3 \gamma _3 \sigma _1
+8 \beta _1 \gamma _1 \sigma _2
+12 \beta _3 \gamma _3 \sigma _2
+\beta _3 \big(\gamma _2
-23 \gamma _1\big) \sigma _3
+\big(\beta _1
-11 \beta _2\big) \gamma _3 \sigma _3\big)
+\alpha _1 \big(
-4 \beta _1 \gamma _2 \sigma _2
-11 \beta _1 \gamma _3 \sigma _3
+\beta _2 \big(8 \gamma _2 \sigma _1
+\gamma _3 \sigma _3\big)
+\beta _3 \big(12 \gamma _3 \sigma _1
-\gamma _3 \sigma _2
+\gamma _1 \sigma _3
-23 \gamma _2 \sigma _3\big)\big)
+\alpha _3 \big(\beta _2 \big(
-23 \gamma _3 \sigma _1
+\gamma _3 \sigma _2
-\gamma _1 \sigma _3
+12 \gamma _2 \sigma _3\big)
+\beta _3 \big(\gamma _2 \sigma _1
-11 \gamma _2 \sigma _2
+\gamma _1 \big(\sigma _2
-11 \sigma _1\big)
+36 \gamma _3 \sigma _3\big)
+\beta _1 \big(\gamma _3 \sigma _1
-23 \gamma _3 \sigma _2
+12 \gamma _1 \sigma _3
-\gamma _2 \sigma _3\big)\big)\big) s
+t^2 \big(\alpha _1 \big(
-8 \beta _1 \gamma _2 \sigma _2
-9 \beta _1 \gamma _3 \sigma _3
+\beta _2 \big(4 \gamma _2 \sigma _1
+4 \gamma _1 \sigma _2
+\gamma _3 \sigma _3\big)
+\beta _3 \big(7 \gamma _3 \sigma _1
-11 \gamma _3 \sigma _2
+7 \gamma _1 \sigma _3
-11 \gamma _2 \sigma _3\big)\big)
+\alpha _2 \big(\beta _3 \big(
-11 \gamma _3 \sigma _1
+7 \gamma _3 \sigma _2
-11 \gamma _1 \sigma _3
+7 \gamma _2 \sigma _3\big)
+\beta _1 \big(4 \gamma _2 \sigma _1
+4 \gamma _1 \sigma _2
+\gamma _3 \sigma _3\big)
-\beta _2 \big(8 \gamma _1 \sigma _1
+9 \gamma _3 \sigma _3\big)\big)
+\alpha _3 \big(\beta _2 \big(
-11 \gamma _3 \sigma _1
+7 \gamma _3 \sigma _2
-11 \gamma _1 \sigma _3
+7 \gamma _2 \sigma _3\big)
+\beta _3 \big(\gamma _2 \sigma _1
-9 \gamma _2 \sigma _2
+\gamma _1 \big(\sigma _2
-9 \sigma _1\big)
+32 \gamma _3 \sigma _3\big)
+\beta _1 \big(7 \gamma _3 \sigma _1
-11 \gamma _3 \sigma _2
+7 \gamma _1 \sigma _3
-11 \gamma _2 \sigma _3\big)\big)\big)\big) m^2
+s t \sqrt{2} \sqrt{s t u} \big(2 \alpha _1 \beta _1 \big(\gamma _3 \big(\sigma _1
-3 \sigma _2\big)
+\big(\gamma _1
-3 \gamma _2\big) \sigma _3\big)
+\alpha _1 \beta _3 \big(
-2 \gamma _1 \sigma _1
-6 \gamma _2 \sigma _2
+7 \gamma _3 \sigma _3\big)
+\alpha _3 \big(
-2 \beta _1 \gamma _1 \sigma _1
+6 \beta _2 \gamma _1 \sigma _1
-7 \beta _3 \gamma _3 \sigma _1
-6 \beta _1 \gamma _2 \sigma _2
+2 \beta _2 \gamma _2 \sigma _2
+7 \beta _3 \gamma _3 \sigma _2
+7 \big(\beta _3 \big(\gamma _2
-\gamma _1\big)
+\big(\beta _1
-\beta _2\big) \gamma _3\big) \sigma _3\big)
+2 \alpha _2 \beta _2 \big(3 \gamma _3 \sigma _1
-\gamma _3 \sigma _2
+3 \gamma _1 \sigma _3
-\gamma _2 \sigma _3\big)
+\alpha _2 \beta _3 \big(6 \gamma _1 \sigma _1
+2 \gamma _2 \sigma _2
-7 \gamma _3 \sigma _3\big)\big) m
+s^2 t \big(2 s \big(
-\beta _3 \gamma _3 \big(\alpha _1 \sigma _1
+\alpha _2 \sigma _2\big)
+\big(2 \alpha _2 \beta _3 \gamma _1
+2 \alpha _1 \beta _3 \gamma _2
+\alpha _1 \beta _1 \gamma _3
+\alpha _2 \beta _2 \gamma _3\big) \sigma _3
+\alpha _3 \big(\beta _3 \gamma _1 \sigma _1
+2 \beta _2 \gamma _3 \sigma _1
+\beta _3 \gamma _2 \sigma _2
+2 \beta _1 \gamma _3 \sigma _2
-\big(\beta _1 \gamma _1
+\beta _2 \gamma _2
+3 \beta _3 \gamma _3\big) \sigma _3\big)\big)
+t \big(
-2 \beta _3 \gamma _3 \big(\alpha _2 \big(\sigma _2
-2 \sigma _1\big)
+\alpha _1 \big(\sigma _1
-2 \sigma _2\big)\big)
-2 \beta _3 \big(\alpha _2 \big(\gamma _2
-2 \gamma _1\big)
+\alpha _1 \big(\gamma _1
-2 \gamma _2\big)\big) \sigma _3
+4 \big(\alpha _1 \beta _1
+\alpha _2 \beta _2\big) \gamma _3 \sigma _3
+\alpha _3 \big(
-2 \gamma _3 \big(\beta _1 \sigma _1
-2 \beta _2 \sigma _1
-2 \beta _1 \sigma _2
+\beta _2 \sigma _2\big)
-2 \big(\beta _1 \gamma _1
-2 \beta _2 \gamma _1
-2 \beta _1 \gamma _2
+\beta _2 \gamma _2\big) \sigma _3
+\beta _3 \big(4 \gamma _1 \sigma _1
+4 \gamma _2 \sigma _2
-11 \gamma _3 \sigma _3\big)\big)\big)\big)\Big)

\end{autobreak}
\end{align}

\begin{align}\label{eq: nuhat}
\begin{autobreak}
   \hat n_{u}=
-\frac{i}{4 \left(s-4 m^2\right)^2} 
\Big(192 \big(\alpha _3 \big(
-\beta _3 \gamma _1 \sigma _1
+\beta _2 \gamma _3 \sigma _1
-\beta _3 \gamma _2 \sigma _2
+\beta _1 \gamma _3 \sigma _2\big)
+\alpha _1 \big(\beta _1 \gamma _1 \sigma _1
-\beta _2 \gamma _2 \sigma _1
+\beta _3 \gamma _2 \sigma _3
-\beta _1 \gamma _3 \sigma _3\big)
+\alpha _2 \big(
-\beta _1 \gamma _1 \sigma _2
+\beta _2 \gamma _2 \sigma _2
+\beta _3 \gamma _1 \sigma _3
-\beta _2 \gamma _3 \sigma _3\big)\big) m^8
-16 \big(2 s \big(\alpha _2 \big(
-7 \beta _1 \gamma _1 \sigma _2
+3 \beta _2 \gamma _2 \sigma _2
-6 \beta _3 \gamma _3 \sigma _2
+13 \beta _3 \gamma _1 \sigma _3\big)
+\alpha _1 \big(3 \beta _1 \gamma _1 \sigma _1
-7 \beta _2 \gamma _2 \sigma _1
-6 \beta _3 \gamma _3 \sigma _1
+13 \beta _3 \gamma _2 \sigma _3\big)
+\alpha _3 \big(13 \beta _2 \gamma _3 \sigma _1
+13 \beta _1 \gamma _3 \sigma _2
-6 \beta _1 \gamma _1 \sigma _3
-6 \beta _2 \gamma _2 \sigma _3
-12 \beta _3 \gamma _3 \sigma _3\big)\big)
+t \big(\alpha _3 \big(\beta _2 \big(
-\gamma _3 \sigma _1
+5 \gamma _3 \sigma _2
-8 \gamma _1 \sigma _3
+4 \gamma _2 \sigma _3\big)
+4 \beta _3 \big(
-3 \gamma _1 \sigma _1
+2 \gamma _2 \sigma _1
+2 \gamma _1 \sigma _2
-3 \gamma _2 \sigma _2
+\gamma _3 \sigma _3\big)
+\beta _1 \big(5 \gamma _3 \sigma _1
-\gamma _3 \sigma _2
+4 \gamma _1 \sigma _3
-8 \gamma _2 \sigma _3\big)\big)
-\alpha _2 \big(\beta _2 \big(
-26 \gamma _1 \sigma _1
+5 \gamma _2 \sigma _1
+5 \gamma _1 \sigma _2
-8 \gamma _2 \sigma _2
+12 \gamma _3 \sigma _3\big)
+\beta _3 \big(8 \gamma _3 \sigma _1
-4 \gamma _3 \sigma _2
+\gamma _1 \sigma _3
-5 \gamma _2 \sigma _3\big)
+\beta _1 \big(5 \gamma _1 \sigma _1
+6 \gamma _1 \sigma _2
+5 \gamma _2 \sigma _2
-8 \gamma _3 \sigma _3\big)\big)
-\alpha _1 \big(\beta _3 \big(
-4 \gamma _3 \sigma _1
+8 \gamma _3 \sigma _2
-5 \gamma _1 \sigma _3
+\gamma _2 \sigma _3\big)
+\beta _1 \big(
-8 \gamma _1 \sigma _1
+5 \gamma _2 \sigma _1
+5 \gamma _1 \sigma _2
-26 \gamma _2 \sigma _2
+12 \gamma _3 \sigma _3\big)
+\beta _2 \big(5 \gamma _1 \sigma _1
+6 \gamma _2 \sigma _1
+5 \gamma _2 \sigma _2
-8 \gamma _3 \sigma _3\big)\big)\big)\big) m^6
-8 \sqrt{2} \sqrt{s t u} \big(\alpha _3 \big(\beta _2 \big(
-13 \gamma _1 \sigma _1
+5 \gamma _2 \sigma _1
-8 \gamma _2 \sigma _2
+19 \gamma _3 \sigma _3\big)
+19 \beta _3 \big(\gamma _3 \sigma _1
-\gamma _3 \sigma _2
+\gamma _1 \sigma _3
-\gamma _2 \sigma _3\big)
+\beta _1 \big(8 \gamma _1 \sigma _1
-5 \gamma _1 \sigma _2
+13 \gamma _2 \sigma _2
-19 \gamma _3 \sigma _3\big)\big)
+\alpha _2 \big(
-5 \beta _1 \gamma _3 \sigma _2
+5 \beta _1 \gamma _1 \sigma _3
+\beta _2 \big(
-13 \gamma _3 \sigma _1
+8 \gamma _3 \sigma _2
-13 \gamma _1 \sigma _3
+8 \gamma _2 \sigma _3\big)
+\beta _3 \big(
-13 \gamma _1 \sigma _1
+5 \gamma _1 \sigma _2
-8 \gamma _2 \sigma _2
+19 \gamma _3 \sigma _3\big)\big)
+\alpha _1 \big(\beta _1 \big(
-8 \gamma _3 \sigma _1
+13 \gamma _3 \sigma _2
-8 \gamma _1 \sigma _3
+13 \gamma _2 \sigma _3\big)
+5 \beta _2 \big(\gamma _3 \sigma _1
-\gamma _2 \sigma _3\big)
+\beta _3 \big(8 \gamma _1 \sigma _1
-5 \gamma _2 \sigma _1
+13 \gamma _2 \sigma _2
-19 \gamma _3 \sigma _3\big)\big)\big) m^5
+4 \big(\big(\alpha _1 \big(3 \beta _1 \gamma _1 \sigma _1
-19 \beta _2 \gamma _2 \sigma _1
-32 \beta _3 \gamma _3 \sigma _1
+59 \beta _3 \gamma _2 \sigma _3
+17 \beta _1 \gamma _3 \sigma _3\big)
+\alpha _2 \big(
-19 \beta _1 \gamma _1 \sigma _2
+3 \beta _2 \gamma _2 \sigma _2
-32 \beta _3 \gamma _3 \sigma _2
+59 \beta _3 \gamma _1 \sigma _3
+17 \beta _2 \gamma _3 \sigma _3\big)
+\alpha _3 \big(59 \beta _2 \gamma _3 \sigma _1
+59 \beta _1 \gamma _3 \sigma _2
-32 \beta _1 \gamma _1 \sigma _3
-32 \beta _2 \gamma _2 \sigma _3
+17 \beta _3 \big(\gamma _1 \sigma _1
+\gamma _2 \sigma _2
-4 \gamma _3 \sigma _3\big)\big)\big) s^2
+t \big(\alpha _2 \big(\beta _2 \big(50 \gamma _1 \sigma _1
-5 \gamma _2 \sigma _1
-5 \gamma _1 \sigma _2
+8 \gamma _2 \sigma _2
+10 \gamma _3 \sigma _3\big)
+\beta _3 \big(6 \gamma _3 \sigma _1
-22 \gamma _3 \sigma _2
+41 \gamma _1 \sigma _3
-7 \gamma _2 \sigma _3\big)
-\beta _1 \big(5 \gamma _1 \sigma _1
+22 \gamma _1 \sigma _2
+5 \gamma _2 \sigma _2
-6 \gamma _3 \sigma _3\big)\big)
+\alpha _1 \big(\beta _3 \big(
-22 \gamma _3 \sigma _1
+6 \gamma _3 \sigma _2
-7 \gamma _1 \sigma _3
+41 \gamma _2 \sigma _3\big)
+\beta _1 \big(8 \gamma _1 \sigma _1
-5 \gamma _2 \sigma _1
-5 \gamma _1 \sigma _2
+50 \gamma _2 \sigma _2
+10 \gamma _3 \sigma _3\big)
-\beta _2 \big(5 \gamma _1 \sigma _1
+22 \gamma _2 \sigma _1
+5 \gamma _2 \sigma _2
-6 \gamma _3 \sigma _3\big)\big)
+\alpha _3 \big(\beta _1 \big(
-7 \gamma _3 \sigma _1
+41 \gamma _3 \sigma _2
-22 \gamma _1 \sigma _3
+6 \gamma _2 \sigma _3\big)
+\beta _2 \big(41 \gamma _3 \sigma _1
-7 \gamma _3 \sigma _2
+6 \gamma _1 \sigma _3
-22 \gamma _2 \sigma _3\big)
+2 \beta _3 \big(5 \gamma _1 \sigma _1
+3 \gamma _2 \sigma _1
+3 \gamma _1 \sigma _2
+5 \gamma _2 \sigma _2
-38 \gamma _3 \sigma _3\big)\big)\big) s
+t^2 \big(4 \alpha _3 \big(\big(\beta _1
-\beta _2\big) \big(\gamma _3 \big(\sigma _1
-\sigma _2\big)
+\big(\gamma _1
-\gamma _2\big) \sigma _3\big)
-2 \beta _3 \big(\gamma _1
-\gamma _2\big) \big(\sigma _1
-\sigma _2\big)\big)
-\alpha _2 \big(\beta _2 \big(
-21 \gamma _1 \sigma _1
+5 \gamma _2 \sigma _1
+5 \gamma _1 \sigma _2
-5 \gamma _2 \sigma _2
+8 \gamma _3 \sigma _3\big)
+4 \beta _3 \big(\gamma _3 \sigma _1
-\gamma _3 \sigma _2
+\gamma _1 \sigma _3
-\gamma _2 \sigma _3\big)
+\beta _1 \big(5 \gamma _1 \sigma _1
+3 \gamma _2 \sigma _1
+3 \gamma _1 \sigma _2
+5 \gamma _2 \sigma _2
-8 \gamma _3 \sigma _3\big)\big)
-\alpha _1 \big(4 \beta _3 \big(
-\gamma _3 \sigma _1
+\gamma _3 \sigma _2
-\gamma _1 \sigma _3
+\gamma _2 \sigma _3\big)
+\beta _1 \big(
-5 \gamma _1 \sigma _1
+5 \gamma _2 \sigma _1
+5 \gamma _1 \sigma _2
-21 \gamma _2 \sigma _2
+8 \gamma _3 \sigma _3\big)
+\beta _2 \big(5 \gamma _1 \sigma _1
+3 \gamma _2 \sigma _1
+3 \gamma _1 \sigma _2
+5 \gamma _2 \sigma _2
-8 \gamma _3 \sigma _3\big)\big)\big)\big) m^4
+2 \sqrt{2} \sqrt{s t u} \big(t \big(\alpha _3 \big(\beta _2 \big(
-17 \gamma _1 \sigma _1
+5 \gamma _2 \sigma _1
+5 \gamma _1 \sigma _2
-9 \gamma _2 \sigma _2
+24 \gamma _3 \sigma _3\big)
+24 \beta _3 \big(\gamma _3 \sigma _1
-\gamma _3 \sigma _2
+\gamma _1 \sigma _3
-\gamma _2 \sigma _3\big)
+\beta _1 \big(9 \gamma _1 \sigma _1
-5 \gamma _2 \sigma _1
-5 \gamma _1 \sigma _2
+17 \gamma _2 \sigma _2
-24 \gamma _3 \sigma _3\big)\big)
+\alpha _2 \big(\beta _2 \big(
-17 \gamma _3 \sigma _1
+9 \gamma _3 \sigma _2
-17 \gamma _1 \sigma _3
+9 \gamma _2 \sigma _3\big)
+\beta _3 \big(
-17 \gamma _1 \sigma _1
+5 \gamma _2 \sigma _1
+5 \gamma _1 \sigma _2
-9 \gamma _2 \sigma _2
+24 \gamma _3 \sigma _3\big)
+5 \beta _1 \big(\gamma _3 \sigma _1
-\gamma _3 \sigma _2
+\gamma _1 \sigma _3
-\gamma _2 \sigma _3\big)\big)
+\alpha _1 \big(\beta _1 \big(
-9 \gamma _3 \sigma _1
+17 \gamma _3 \sigma _2
-9 \gamma _1 \sigma _3
+17 \gamma _2 \sigma _3\big)
+5 \beta _2 \big(\gamma _3 \sigma _1
-\gamma _3 \sigma _2
+\gamma _1 \sigma _3
-\gamma _2 \sigma _3\big)
+\beta _3 \big(9 \gamma _1 \sigma _1
-5 \gamma _2 \sigma _1
-5 \gamma _1 \sigma _2
+17 \gamma _2 \sigma _2
-24 \gamma _3 \sigma _3\big)\big)\big)
+s \big(\alpha _3 \big(\beta _2 \big(
-25 \gamma _1 \sigma _1
+5 \gamma _2 \sigma _1
-12 \gamma _2 \sigma _2
+33 \gamma _3 \sigma _3\big)
+33 \beta _3 \big(\gamma _3 \sigma _1
-\gamma _3 \sigma _2
+\gamma _1 \sigma _3
-\gamma _2 \sigma _3\big)
+\beta _1 \big(12 \gamma _1 \sigma _1
-5 \gamma _1 \sigma _2
+25 \gamma _2 \sigma _2
-33 \gamma _3 \sigma _3\big)\big)
+\alpha _2 \big(
-5 \beta _1 \gamma _3 \sigma _2
+5 \beta _1 \gamma _1 \sigma _3
+\beta _2 \big(
-25 \gamma _3 \sigma _1
+12 \gamma _3 \sigma _2
-25 \gamma _1 \sigma _3
+12 \gamma _2 \sigma _3\big)
+\beta _3 \big(
-12 \gamma _2 \sigma _2
+5 \gamma _1 \big(\sigma _2
-5 \sigma _1\big)
+33 \gamma _3 \sigma _3\big)\big)
+\alpha _1 \big(\beta _1 \big(
-12 \gamma _3 \sigma _1
+25 \gamma _3 \sigma _2
-12 \gamma _1 \sigma _3
+25 \gamma _2 \sigma _3\big)
+5 \beta _2 \big(\gamma _3 \sigma _1
-\gamma _2 \sigma _3\big)
+\beta _3 \big(12 \gamma _1 \sigma _1
-5 \gamma _2 \sigma _1
+25 \gamma _2 \sigma _2
-33 \gamma _3 \sigma _3\big)\big)\big)\big) m^3
-2 s \big(\big(\alpha _1 \big(
-4 \beta _2 \gamma _2 \sigma _1
-14 \beta _3 \gamma _3 \sigma _1
+26 \beta _3 \gamma _2 \sigma _3
+11 \beta _1 \gamma _3 \sigma _3\big)
+\alpha _2 \big(
-4 \beta _1 \gamma _1 \sigma _2
-14 \beta _3 \gamma _3 \sigma _2
+26 \beta _3 \gamma _1 \sigma _3
+11 \beta _2 \gamma _3 \sigma _3\big)
+\alpha _3 \big(11 \beta _3 \gamma _1 \sigma _1
+26 \beta _2 \gamma _3 \sigma _1
+11 \beta _3 \gamma _2 \sigma _2
+26 \beta _1 \gamma _3 \sigma _2
-14 \beta _1 \gamma _1 \sigma _3
-14 \beta _2 \gamma _2 \sigma _3
-32 \beta _3 \gamma _3 \sigma _3\big)\big) s^2
+t \big(\alpha _2 \big(
-\beta _1 \big(8 \gamma _1 \sigma _2
+\gamma _3 \sigma _3\big)
+\beta _2 \big(12 \gamma _1 \sigma _1
+23 \gamma _3 \sigma _3\big)
+\beta _3 \big(15 \gamma _3 \sigma _1
-21 \gamma _3 \sigma _2
+37 \gamma _1 \sigma _3
-10 \gamma _2 \sigma _3\big)\big)
+\alpha _1 \big(\beta _3 \big(
-21 \gamma _3 \sigma _1
+15 \gamma _3 \sigma _2
-10 \gamma _1 \sigma _3
+37 \gamma _2 \sigma _3\big)
-\beta _2 \big(8 \gamma _2 \sigma _1
+\gamma _3 \sigma _3\big)
+\beta _1 \big(12 \gamma _2 \sigma _2
+23 \gamma _3 \sigma _3\big)\big)
+\alpha _3 \big(\beta _1 \big(
-10 \gamma _3 \sigma _1
+37 \gamma _3 \sigma _2
-21 \gamma _1 \sigma _3
+15 \gamma _2 \sigma _3\big)
+\beta _2 \big(37 \gamma _3 \sigma _1
-10 \gamma _3 \sigma _2
+15 \gamma _1 \sigma _3
-21 \gamma _2 \sigma _3\big)
+\beta _3 \big(23 \gamma _1 \sigma _1
-\gamma _2 \sigma _1
-\gamma _1 \sigma _2
+23 \gamma _2 \sigma _2
-72 \gamma _3 \sigma _3\big)\big)\big) s
+t^2 \big(\alpha _2 \big(
-\beta _1 \big(4 \gamma _2 \sigma _1
+4 \gamma _1 \sigma _2
+\gamma _3 \sigma _3\big)
+\beta _2 \big(8 \gamma _1 \sigma _1
+9 \gamma _3 \sigma _3\big)
+\beta _3 \big(11 \gamma _3 \sigma _1
-7 \gamma _3 \sigma _2
+11 \gamma _1 \sigma _3
-7 \gamma _2 \sigma _3\big)\big)
-\alpha _1 \big(\beta _2 \big(4 \gamma _2 \sigma _1
+4 \gamma _1 \sigma _2
+\gamma _3 \sigma _3\big)
-\beta _1 \big(8 \gamma _2 \sigma _2
+9 \gamma _3 \sigma _3\big)
+\beta _3 \big(7 \gamma _3 \sigma _1
-11 \gamma _3 \sigma _2
+7 \gamma _1 \sigma _3
-11 \gamma _2 \sigma _3\big)\big)
+\alpha _3 \big(\beta _1 \big(
-7 \gamma _3 \sigma _1
+11 \gamma _3 \sigma _2
-7 \gamma _1 \sigma _3
+11 \gamma _2 \sigma _3\big)
+\beta _2 \big(11 \gamma _3 \sigma _1
-7 \gamma _3 \sigma _2
+11 \gamma _1 \sigma _3
-7 \gamma _2 \sigma _3\big)
+\beta _3 \big(9 \gamma _1 \sigma _1
-\gamma _2 \sigma _1
-\gamma _1 \sigma _2
+9 \gamma _2 \sigma _2
-32 \gamma _3 \sigma _3\big)\big)\big)\big) m^2
-\sqrt{2} s (s
+t) \sqrt{s t u} \big(2 \alpha _2 \beta _2 \big(\gamma _3 \big(\sigma _2
-3 \sigma _1\big)
+\big(\gamma _2
-3 \gamma _1\big) \sigma _3\big)
-2 \alpha _1 \beta _1 \big(\gamma _3 \big(\sigma _1
-3 \sigma _2\big)
+\big(\gamma _1
-3 \gamma _2\big) \sigma _3\big)
+\alpha _2 \beta _3 \big(
-6 \gamma _1 \sigma _1
-2 \gamma _2 \sigma _2
+7 \gamma _3 \sigma _3\big)
+\alpha _1 \beta _3 \big(2 \gamma _1 \sigma _1
+6 \gamma _2 \sigma _2
-7 \gamma _3 \sigma _3\big)
+\alpha _3 \big(\beta _2 \big(
-6 \gamma _1 \sigma _1
-2 \gamma _2 \sigma _2
+7 \gamma _3 \sigma _3\big)
+7 \beta _3 \big(\gamma _3 \sigma _1
-\gamma _3 \sigma _2
+\gamma _1 \sigma _3
-\gamma _2 \sigma _3\big)
+\beta _1 \big(2 \gamma _1 \sigma _1
+6 \gamma _2 \sigma _2
-7 \gamma _3 \sigma _3\big)\big)\big) m
+s^2 (s
+t) \big(s \big(2 \big(\alpha _1 \big(
-\beta _3 \gamma _3 \sigma _1
+2 \beta _3 \gamma _2 \sigma _3
+\beta _1 \gamma _3 \sigma _3\big)
+\alpha _2 \big(
-\beta _3 \gamma _3 \sigma _2
+2 \beta _3 \gamma _1 \sigma _3
+\beta _2 \gamma _3 \sigma _3\big)\big)
+\alpha _3 \big(4 \beta _2 \gamma _3 \sigma _1
+4 \beta _1 \gamma _3 \sigma _2
-2 \beta _1 \gamma _1 \sigma _3
-2 \beta _2 \gamma _2 \sigma _3
+\beta _3 \big(2 \gamma _1 \sigma _1
+2 \gamma _2 \sigma _2
-5 \gamma _3 \sigma _3\big)\big)\big)
+t \big(4 \alpha _1 \beta _1 \gamma _3 \sigma _3
+4 \alpha _2 \beta _2 \gamma _3 \sigma _3
-2 \alpha _1 \beta _3 \big(\gamma _3 \big(\sigma _1
-2 \sigma _2\big)
+\big(\gamma _1
-2 \gamma _2\big) \sigma _3\big)
+2 \alpha _2 \beta _3 \big(2 \gamma _3 \sigma _1
-\gamma _3 \sigma _2
+2 \gamma _1 \sigma _3
-\gamma _2 \sigma _3\big)
+\alpha _3 \big(\beta _3 \big(4 \gamma _1 \sigma _1
+4 \gamma _2 \sigma _2
-11 \gamma _3 \sigma _3\big)
-2 \big(\beta _2 \gamma _3 \big(\sigma _2
-2 \sigma _1\big)
+\beta _1 \gamma _3 \big(\sigma _1
-2 \sigma _2\big)
+\beta _2 \big(\gamma _2
-2 \gamma _1\big) \sigma _3
+\beta _1 \big(\gamma _1
-2 \gamma _2\big) \sigma _3\big)\big)\big)\big)\Big)

\end{autobreak}
\end{align}

%%%%%%%%%%%%%%%%
\bibliographystyle{JHEP}
\bibliography{references}
%%%%%%%%%%%%%%%%

\end{document}